\documentclass[reprint,nofootinbib,showpacs,showkeys,prd]{revtex4-1}
\usepackage{graphicx}
\usepackage{epsfig}
\usepackage{amsmath}
\usepackage{amssymb}
\usepackage{calc}
\usepackage{verbatim}
\usepackage[caption=false]{subfig}
\usepackage[unicode]{hyperref}
\begin{document}
\title{Cluster algebraic description of entanglement patterns for the BTZ black hole}
\author{Bercel Boldis and P\'eter L\'evay}
\affiliation{MTA-BME Quantum Dynamics and Correlations Research Group, Department of Theoretical Physics,
Budapest University of  Technology and Economics, 1521 Budapest, Hungary}

\date{\today}
\begin{abstract}
We study the thermal state of a two dimensional conformal field theory which is dual to the static BTZ black hole in the high temperature limit.
After partitioning the boundary of the static BTZ slice into $N$ subsystems we show that there is an underlying $C_{N-1}$ cluster algebra encoding entanglement patterns of the thermal state. We also demonstrate that the polytope encapsulating such patterns in a geometric manner for a fixed $N$ is the cyclohedron ${\mathcal C}_{N-1}$.
Alternatively these patterns of entanglement can be represented in the space of geodesics (kinematic space) in terms of a Zamolodchikov $Y$-system of $C_{N-1}$ type. The boundary condition for such an $Y$-system is featuring the entropy of the BTZ black hole. 
\end{abstract}

\pacs{}
\keywords{${\rm AdS}_3/{\rm CFT}_2$ correspondence, Cluster algebras,
   Kinematic Space, Quantum Entanglement}

\maketitle
\section{Introduction}

Since the advent of the Ryu-Takayanagi formula\cite{RT,RT2,HRT} it has become clear that the entanglement structure of certain quantum states of a $d$-dimensional boundary Conformal Field Theory (CFT) is encoded into classical geometric structures of a $d+1$ dimensional bulk.
The role of the classical geometric structures is played by extremal surfaces in the bulk homologous to boundary regions with their associated reduced density matrices encapsulating entanglement information.
It turned out that in order to obtain a deeper understanding of this encoding it is rewarding to consider the space of these extremal surfaces. This space as a geometric entity is playing the role of an intermediary between the spaces of the bulk and the boundary\cite{Czech1,Czech1b}.
For example for the  $d=2$ case one is considering the $AdS_3/CFT_2$ correspondence where the extremal surfaces are geodesics of the asymptotically $AdS_3$ geometry. This space of geodesics is the kinematic space\cite{Czech1,Czech1b}  $\mathbb K$.
An alternative characterization of $\mathbb K$ as the the moduli space of boundary causal diamonds has also appeared\cite{Myers}.

The advantage of the introduction of these spaces rests in the possibility of a clear understanding of how patterns of entanglement manifest themselves in patterns of geometry.
For example in the  $d=2$ case the conditional mutual informations of entangled intersecting domains\cite{NC,CS} of the boundary appear in kinematic space as areas of certain regions with respect to an area form called the Crofton form\cite{Czech1}.
In particular it is known that taking the static slice of $AdS_3$ entanglement quantities of the CFT vacuum are represented by area labels of causal diamonds in a 1+1 dimensional se Sitter ($dS_2$) space\cite{Czech1} playing the role of kinematic space.

However, much more can be revealed.
In our previous paper\cite{LB} in the simplest case of pure $AdS_3$ dual to the vacuum state of a $CFT_2$ we elaborated on this representation.
We have shown, that area labels of $dS_2$ causal diamonds are encoding entanglement information via Zamolodchikov $Y$-systems\cite{Zamo} well-known from studies of integrable systems\cite{FrenkelSzenes,Ravanini,Gliozzi}.

In arriving at this result another space, the space of horocycles related to the gauge degree of freedom\cite{Czech2} for choosing a cutoff for regularizing the divergent entanglement entropies has also appeared\cite{Levay}.
Horocycles are geometric objects regularizing the diverging length of geodesics. Employing them resulted in the observation that the lambda lengths of Penner\cite{Penner,Pennerbook}, encapsulating this regularization in a geometric manner, are direcly related to entanglement entropies via the Ryu-Takayanagi formula\cite{RT}. 
An extra bonus of this observation was the realization\cite{Levay} that lambda lengths also provide a link to cluster algebras\cite{ClusterFZ,Fomin,Williams}, algebraic structures that are under intense scrutiny in the mathematics literature.

Cluster algebras are defined recursively via certain transformations called flips. The results of Ref.\cite{LB} show that these flips are related to mutations between patterns of entanglement associated to a partition of the boundary to $N$ subsystems. 
For example in the static pure $AdS_3/CFT_2$ scenario one can choose the quantum state associated to the boundary as the $CFT_2$ vacuum. Then one  fixes a partition of the boundary to $N$ subsystems. To this fixed partition there are many geodesic triangulations of the bulk. Flips are operating between such triangulations, by exchanging the two possible diagonals of geodesic quadrangles. Using successive flips in the space of triangulations one can define an $N-3$ dimensional polytope the associahedron ${\mathcal A}_{N-3}$ encapsulating entanglement information in a geometric manner. Moreover, one can show\cite{LB} that such flips define recursively a cluster algebra of type  $A_{N-3}$. Hence one arrives at the result\cite{LB} that after partitioning the static slice of the boundary to $N$ subsystems the entanglement properties of the $CFT_2$ vacuum are encoded into the structure of an $A_{N-3}$ cluster algebra. 

Based on this result one can conjecture that a similar association of entanglement patterns of other $CFT_2$ states and cluster algebras might exist. 
In this paper we show that this is really the case. 
We show that the thermal state of the $CFT_2$ which is dual to the static BTZ black hole\cite{BTZ} in the high temperature limit
provides a nontrivial example of that kind.
We show that after partitioning the static BTZ slice into $N$ subsystems the underlying cluster algebra encoding entanglement patterns is a $C_{N-1}$ one. Moreover, it turns out that in this new case the polytope encapsulating such patterns in a geometric manner for a fixed $N$ is the cyclohedron ${\mathcal C}_{N-1}$.
Displaying these patterns of entanglement in kinematic space reveals that their algebraic structure is connected to a Zamolodchikov $Y$-system of $C_{N-1}$ type. The boundary condition for such an $Y$-system is featuring the entropy of the BTZ black hole showing up in area labels of boundary triangles representing conditional entropies. 
We hope that our results will pave the way for further elaborations exploring the mathematical properties of quantum states
where such cluster algebraic connection shows up.

The organization of this paper is as follows.
In Section II. the basic properties of the BTZ black hole in the high temperature limit are reviewed.
Section III. is devoted to a short reminder on lambda lengths and their connection to geodesics in the BTZ context.
In Section IV. we summarize the basic quantities of quantum information theoretic meaning relevant for our elaborations. Armed with the background material of these sections in Section V. we explore the properties of geodesic triangulations of the bulk displaying the geometric structure of a BTZ black hole. Then we show that for a partition of the boundary featuring $N$ subsystems the associated lambda lengths are generating a $C_{N-1}$ cluster algebra. 
We observe that the exchange graph of this algebra is the cyclohedron ${\mathcal C}_{N-1}$. We illustrate its structure
in the BTZ picture, displaying different types of flips of BTZ geodesics. 
In section VI. we explore how these algebraic structures are represented in the BTZ kinematic space. We find that the set of areas of causal diamonds in the BTZ kinematic space can be expressed in terms of cross ratios, and that they can be organized into an Y-system characterizing the BTZ scenario in the high temperature limit.
The $Y$-system we find is a Zamolodchikov system of type $C_{N-1}$ with a boundary condition featuring the entropy of the BTZ black hole.
Our conclusions and some comments are left for Section VII.

\section{BTZ black hole}

Three dimensional anti de Sitter space AdS\textsubscript{3} is defined as the set of points of the flat $\mathbb{R}^{2,2}$
space satisfying the constraint

\begin{equation}\label{eq:constraint}
-U^2-V^2+X^2+Y^2=-R^2
\end{equation}
where $R$ is the AdS radius. The induced metric of this space is

\begin{equation}\label{eq:metric_ads}
    ds^2=-dU^2-dV^2+dX^2+dY^2
\end{equation}

It is well-known \cite{Brill,Ingemar1, Skenderis} that multiboundary wormhole solutions of Einstein's equations with negative cosmological constant can be obtained from $AdS_3$ by factorizing this space by the action of a suitable discrete subgroup of its isometry group.
In this paper we are focusing on the the simplest solution of that kind, namely the BTZ black hole\cite{BTZ}. It is characterized by two parameters, namely the mass and angular momentum of the black hole. In the special case when the angular momentum is set to zero the solution is given in terms of Schwarzschild coordinates 

\begin{equation}\label{eq:coords}
    \left(\begin{array}{c}
U \\
V \\
X \\
Y
\end{array}\right)=\left(\begin{array}{c}
\frac{R}{r_+}r \cosh \left(\frac{r_+}{R}\varphi\right) \\
\sqrt{\left(\frac{R}{r_+}r\right)^{2}-R^{2}} \sinh \left(\frac{r_+}{R}\tau\right) \\
\frac{R}{r_+}r \sinh \left(\frac{r_+}{R}\varphi \right)\\
\sqrt{\left(\frac{R}{r_+}r\right)^{2}-R^{2}} \cosh \left(\frac{r_+}{R}\tau\right)
\end{array}\right)
\end{equation}
subject to certain constraints coming from this group action.
Namely, the corresponding action\cite{Brill,Ingemar1, Skenderis,Carlip} boils down to a $\varphi\sim\varphi+2\pi$ identification of the hyperbolic angle, so that $-\pi<\varphi<\pi$.

Using \eqref{eq:metric_ads} in terms of these coordinates the BTZ metric becomes

\begin{equation}\label{eq:metric_hyp}
    d s^{2}=-\left(r^{2}-r_{+}^{2}\right) d \tau^{2}+\frac{R^{2}}{r^{2}-r_{+}^{2}} d r^{2}+r^{2} d \varphi^{2}
\end{equation}
where the BTZ black hole solution has an event horizon at $r=r_+$. The boundary of the geometry is given by $r\rightarrow\infty$ and we denote it by $\partial BTZ$.

We can also introduce new scaled coordinates

\begin{equation}
\label{eq:kell}
    \left(\begin{array}{c}
r \\
\tau \\
\varphi
\end{array}\right)=\left(\begin{array}{c}
r_+\cosh{\rho} \\
\frac{R}{r_+}\Theta \\
\frac{R}{r_+}t
\end{array}\right)
\end{equation}
With these variables the metric takes the following form:

\begin{equation}\label{eq:btz2}
    d s^{2}=R^2\left(\cosh^2\rho dt^2+d\rho^2-\sinh^2\rho d\Theta^2\right)
\end{equation}
Now the metric is independent of the size of the horizon $r_+$, but the ranges of $\Theta$ and $t$ are depending on it.

In these considerations the AdS length scale and horizon radius are related to the mass of the black hole via $M=r_+^2/R^2$ (in units where $8G=1$). From now on we deal with the $r_+\gg R$ macroscopic limit of the static slice of the BTZ black hole, that is the mass $M\gg1$. This means that the range of the scaled angle variable is formally $-\infty<t<\infty$ while the periodic identification still holds.
Recall that at high temperature the gravity dual of the $CFT_2$ is the Euclidean version of our BTZ black hole\cite{RT}. Then we have the correspondence $\beta/L=R/r_+\ll1$ where $L=2\pi r_0$ is the systems size with $r_0$ is the cutoff taken to be large and $\beta=1/T$ is the inverse temperature. Hence our macroscopic limit also corresponds to the high temperature limit (HTL).

In the static case we choose $\tau=0$ (i.e. $V=0$ and $\Theta$=0). From \eqref{eq:btz2} the metric of the constant time slice is

\begin{equation}\label{eq:static_btz}
    d s^{2}=R^2\left(\cosh^2\rho dt^2+d\rho^2\right)
\end{equation}
With an alternative set of coordinates we can map this static slice into the Poincar\'e disk $\mathbb{D}$

\begin{equation}\label{eq:D}
    z=\frac{X+i Y}{R+U}=x+i y=\vert z\vert e^{i\vartheta} \in \mathbb{D}
\end{equation}
$\vert z\vert<1$.
In these coordinates the \eqref{eq:metric_ads} metric takes the following form

\begin{equation}\label{eq:metricD}
    ds^2=\frac{4R^2}{(1-x^2-y^2)^2}(dx^2+dy^2)
\end{equation}
The $\partial\mathbb{D}$ boundary of the geometry is obtained by taking the $\vert z\vert ^2=x^2+y^2\rightarrow 1$ limit, yielding the complex unit circle.

\begin{figure}[t]
    \centering
    \includegraphics[width=0.9\columnwidth]{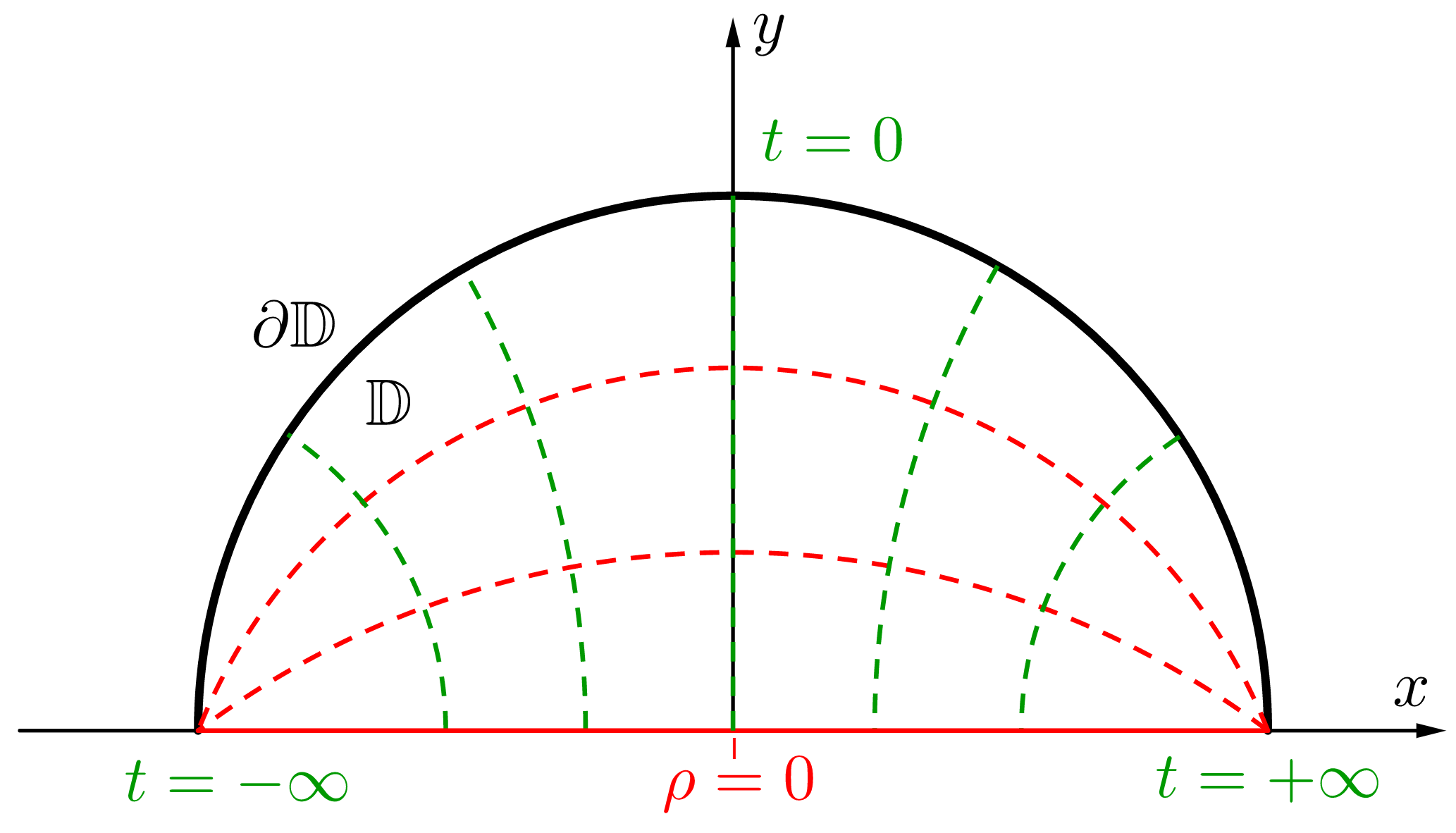}
    \caption{The Poincar\'e disk representation of the static, macroscopic BTZ blackhole. The space covers the upper half of the disk. The red dashed lines correspond to different values of the radial coordinate $\rho$. The $\rho=0$ diameter represents the horizon of the black hole. The green dashed lines are $t=\text{const.}$ lines. The $x=\pm1$ points are identified.}
    \label{fig:BTZ_disk}
\end{figure}

For $\tau=0$ and $-\infty\leq\varphi\leq +\infty$, we obtain the so called BTZ black string. According to \eqref{eq:coords} and \eqref{eq:D} in this case the resulting space is just the upper half of the Poincar\'e disk. If we also make the $\varphi\sim\varphi+2\pi$ identification characterizing our BTZ black hole, then it means that we factorize the disk by a corresponding discrete subgroup. This formally means that we cut out a segment of the upper semi-disk bounded by two geodesics (which are perpendicular to the horizon) and we glue together this segment along these two geodesics. However, in the macroscopic limit these two geodesics shrink to the $\vartheta=0$ and $\vartheta=\pi$ points so the covered segment is the semi-disk itself\cite{Zukowski}, whose $\vartheta=0$ and $\vartheta=\pi$ points are identified (see \hyperref[fig:BTZ_disk]{FIG. 1.}). The $y=0$ diameter is the horizon of the black hole. In this case the conversion between the hyperbolic angle $t$ and the disk angle $\vartheta$ is given by the formula

\begin{equation}\label{eq:angles}
    e^t=\cot{\frac{\vartheta}{2}}
\end{equation}

\section{Geodesics and the lambda length}\label{sec:3}

Now we examine the minimal geodesics between boundary points of the static BTZ black hole in the macroscopic limit. First let us mark two points $a$ and $b$ on the boundary of the BTZ black string\cite{Zukowski} with hyperbolic angles $-\infty<\varphi_a<\varphi_b<\infty$. These two points serve as endpoints of a geodesic.
If we make the identification $\varphi\sim\varphi+2\pi$ to get the static BTZ black hole geometry,
 depending on the mutual positions of $a$ and $b$
we get different types of geodesics. From now on we restrict to geodesics whose endpoints satisfies that $\varphi_b-\varphi_a\leq2\pi$, i.e. they do not intersect with themselves in the bulk. We call them as minimal geodesics. If $\varphi_b-\varphi_a=2\pi$ (so that $a\sim b$), the geodesic winds around the horizon and its endpoints coincide.  We refer to these type of geodesics as loops. Finally if $\varphi_b-\varphi_a<2\pi$, then there are two different arcs between the two marked points. One bypass the hole in one direction and another in the other direction, see FIG 2a.

How can one represent BTZ black hole geodesics on the Poincar\'e disk? First let us describe in general the minimal arcs of the disk geometry. It is well-know that they are given by the following equation\cite{Voros}

\begin{equation}
\left(x-\frac{B_{1}}{M}\right)^{2}+\left(y-\frac{B_{2}}{M}\right)^{2}=\frac{1}{M^{2}}
\end{equation}
where $B_1,B_2$ and $M$ are conserved quantities of the geodesic motion. Hence we see that the geodesics are circular arcs whose endpoints are on the boundary. Let us denote the midpoint coordinate and the half of the opening angle of the boundary interval lying between the two endpoints, by $\theta$ and $\alpha$ respectively. 
Hence we have

\begin{equation}
\label{eq:alfateta}
\theta=(\vartheta_b+\vartheta_a)/2,\qquad
\alpha=(\vartheta_b-\vartheta_a)/2.
\end{equation}

Then we can express the parameters in the geodesic equation by the $\theta$ and $\alpha$ variables as\cite{Czech1}

\begin{equation}\label{eq:conserved}
B_{1}=\frac{\cos \theta}{\sin \alpha} \quad B_{2}=\frac{\sin \theta}{\sin \alpha} \quad M=\frac{\cos \alpha}{\sin \alpha}
\end{equation}

The points terminating the geodesics are of the form $e^{iu},e^{iv}\in\partial\mathbb{D}$. Hence another useful parametrization for geodesics is given by the $(u,v)$ pairs, where $u$ stands for the complex argument of the starting, and $v$ for the ending point. $u$ and $v$ can be expressed by $\theta$ and $\alpha$

\begin{equation}\label{eq:uv}
u=\theta-\alpha \quad v=\theta+\alpha
\end{equation}
The Poincar\'e model, a geodesic and its parameters are shown in \hyperref[fig:disk_geodesic]{FIG. 3.}

\begin{figure*}[t]
    \hspace{0.5cm}
    \subfloat[]{
        \includegraphics[width=0.5\columnwidth]{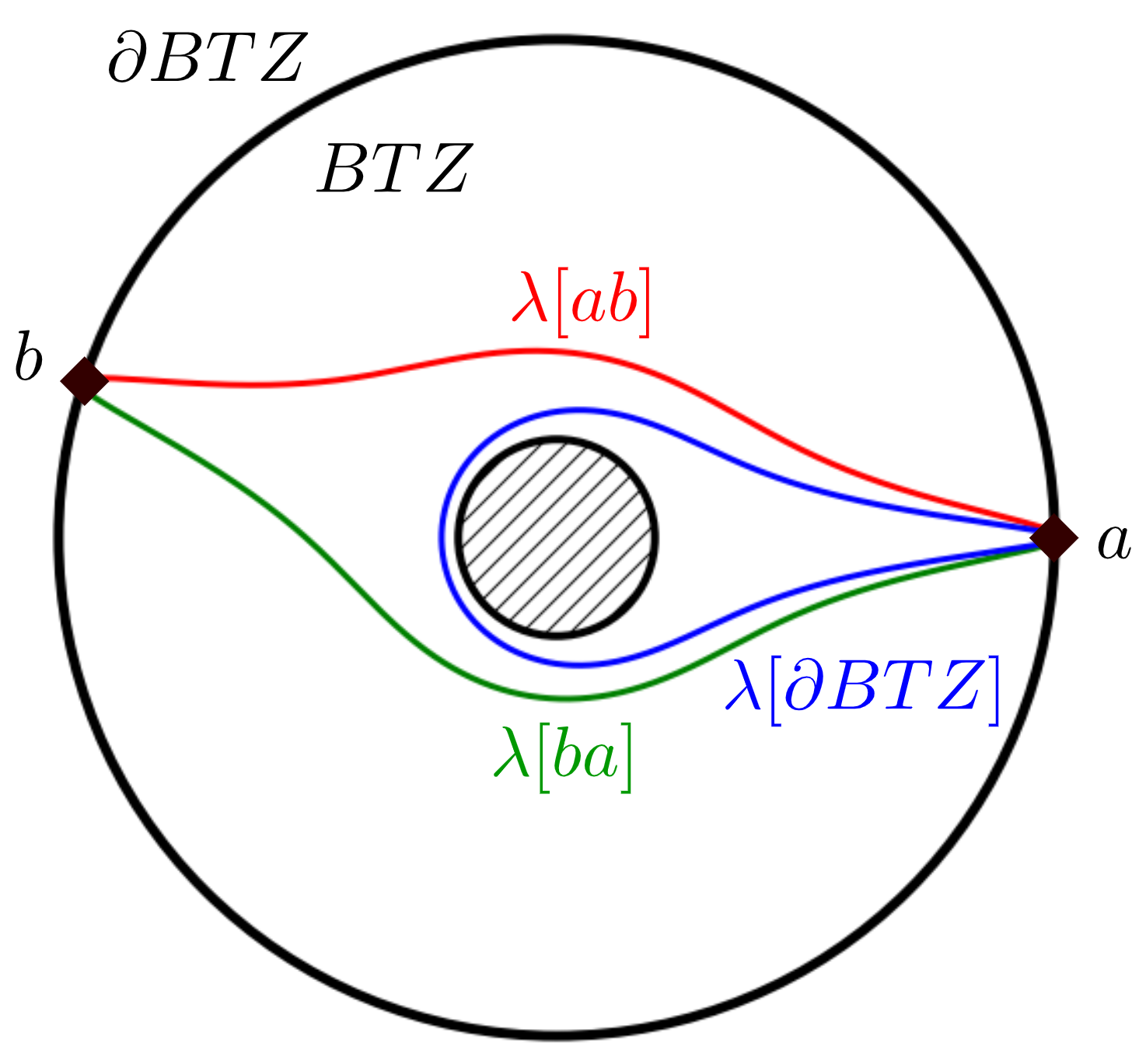}}
    \hfill
    \subfloat[]{
        \includegraphics[width=0.65\columnwidth]{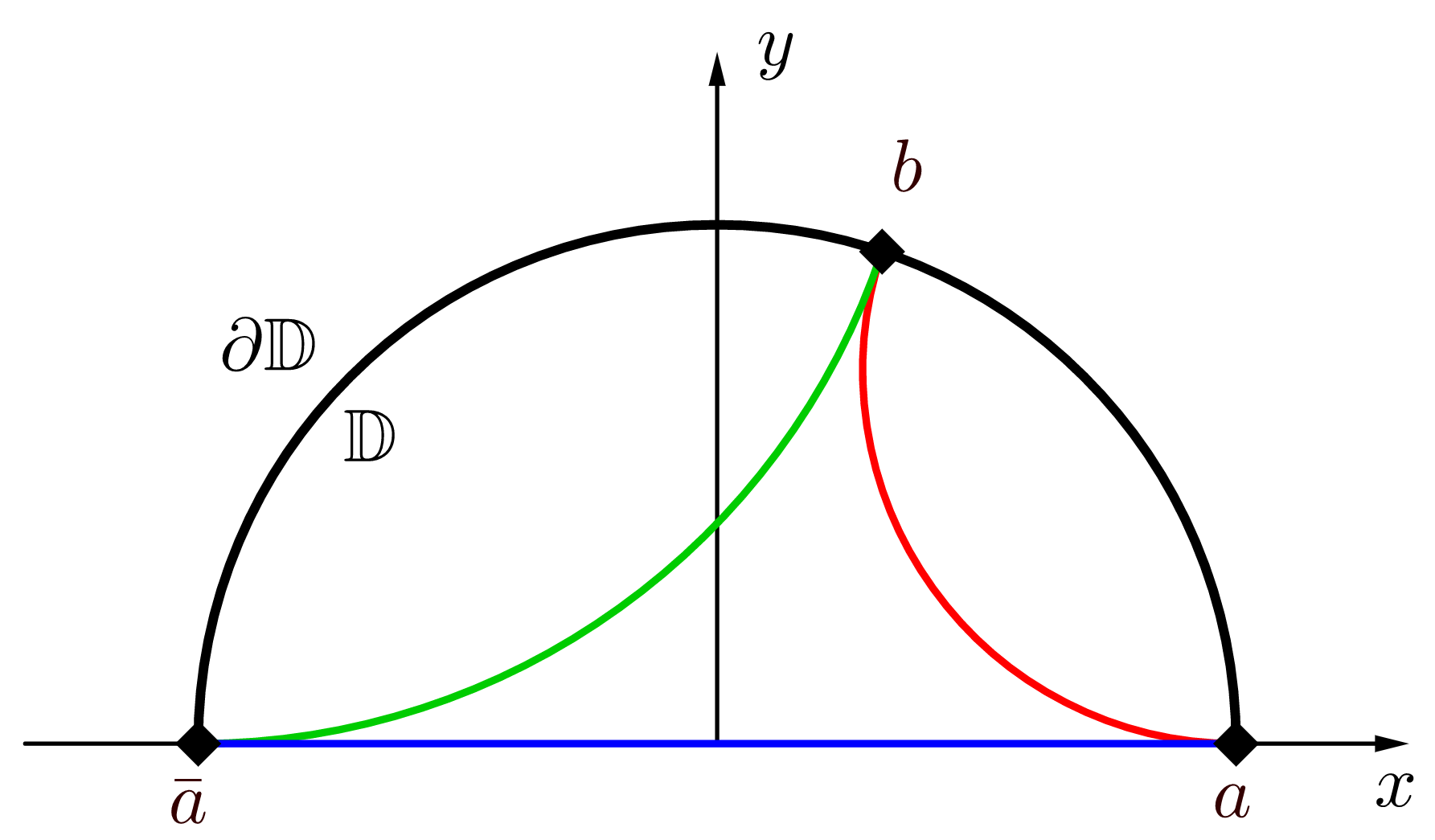}}
    \hfill
    \subfloat[]{
        \includegraphics[width=0.55\columnwidth]{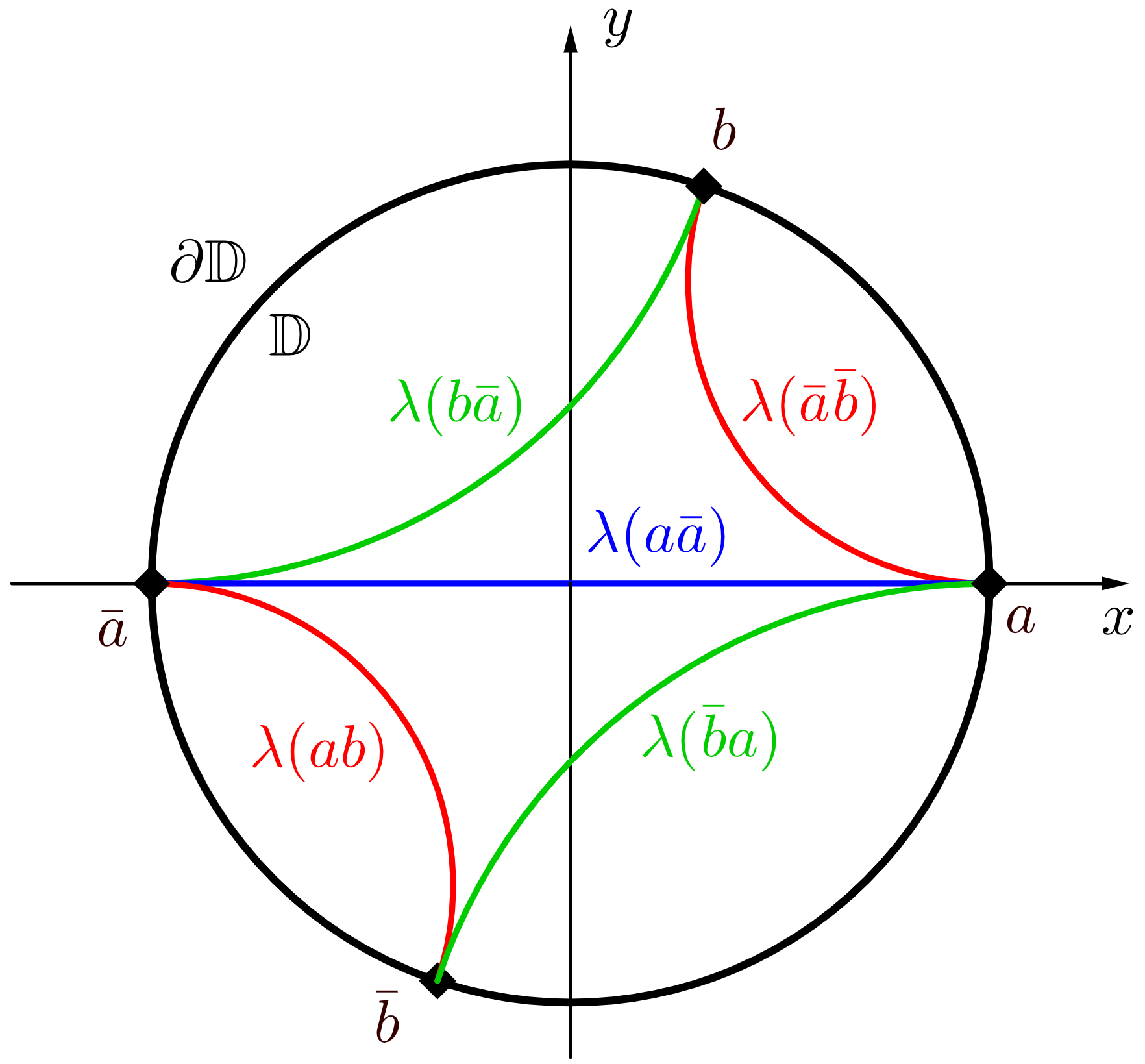}}
    \hspace{0.5cm}
    \caption{Geodesics of the BTZ geometry in the AdS representation (a), and in the disk representation (b). Both the red and green geodesics are connecting the points $a$ and $b$, but they are getting around the horizon on different sides. The blue curve is a loop geodesics, starting and ending in $a$. On the Poincar\'e disk, this can be represented by a diameter between the points $a$ and $\bar{a}\sim a$.
(c) Lambda lengths of a doubly marked BTZ black hole in the disk representation.
}
    \label{fig:geod_representations}
\end{figure*}

Now examine some special cases of black hole geodesics in the disk representation. As we mentioned before, if we transform the static slice of the BTZ black-hole to the Poincar\'e disk, in the macroscopic limit it will cover half of the unit disk, where the $\vartheta=0$ and $\vartheta=\pi$ points will be identified. If we mark a point at $\vartheta=0$ then the loop geodesic that starts and ends at this point will correspond to the $y=0$ diameter of the disk because of the $\vartheta=0\sim\vartheta=\pi$ identification. We can then fix  two different points on the boundary and choose one of them to be the $\vartheta_a=0$ point, and the other one to be arbitrary with coordinate $0<\vartheta_b<\pi$. Again because of the identification, the $a$ point with $\vartheta_a=0$ and the $\bar{a}$ one with $\vartheta_{\bar{a}}=\pi$ represents the same point on the BTZ boundary. Then as in the black hole case, there are two geodesics between the points $a\sim \bar{a}$ and $b$. One of them is going from $\vartheta_a=0$ to $\vartheta_b$ and the other is going from $\vartheta_b$ to $\vartheta_{\bar{a}}=\pi$.

Let us consider a disk geodesic with endpoints $a$ and $b$ with a half opening angle $\alpha$. One can show that its regularized length with respect to the metric \eqref{eq:metricD} is given by\cite{RT,Czech1}

\begin{equation}\label{eq:disk_length}
    \ell(ab)=2R\log\left(e^{\rho_0}\sin\alpha\right)
\end{equation}
Where we used \eqref{eq:alfateta} and regulated the length by restricting the space to the region $x^2+y^2\leq\tanh^2{\rho_0/2}$, assuming that $e^{\rho_0}\gg1$. 

Let us now introduce the following quantity
\begin{equation}\label{eq:lambda_D}
    \lambda(ab)=e^{\ell(ab)/2R}=e^{\rho_0}\sin\alpha
\end{equation}

 For $R=1$ this quantity corresponds to the so called lambda length introduced by Penner\cite{Penner,Pennerbook,Levay}. Using the notion of the lambda length in\cite{Levay} it was shown that it is rewarding to regularize the length of a geodesic by introducing horocycles. A horocycle associated to a boundary point is a circle in the bulk touching the boundary merely at this boundary point. Then the regularized length of a geodesic is that finite length part of it, which is lying entirely outside (or inside) both of the circles associated to the endpoints of the geodesic. It is known that choosing a horocycle associated to a boundary point can be interpreted as a choice of gauge\cite{Czech2,Levay}.  Now the regularization of \eqref{eq:lambda_D} usually showing up in the literature
is a uniform one corresponding to a special uniform choice for the horocycles with their infinitesimally small radii being equal and related to the large value of $\rho_0$.

One can use the notion of a lambda length to identify special classes of geodesics. If we mark a point at $\vartheta=0$, then the loop geodesic that starts and ends at the same BTZ boundary point has got a lambda length

\begin{equation}\label{eq:lambda1}
    \lambda(a\bar{a})=e^{\rho_0}\sin\frac{\pi}{2}
\end{equation}
because this corresponds to the $y=0$ diameter with half opening angle $\alpha=\pi/2$. Notice that this is the longest possible minimal geodesic.
Next we fix two different points on the BTZ boundary. We choose one of them to be the $\vartheta_a=0\sim\vartheta_{\bar{a}}=\pi$ point, and the other to be $0<\vartheta_b<\pi$. Then using
 \eqref{eq:alfateta} the geodesic going from $\vartheta_a=0$ to $\vartheta_b$ has a lambda length

\begin{equation}\label{eq:lambda2}
    \lambda(ab)=e^{\rho_0}\sin\frac{\vartheta_b}{2}
\end{equation}
and the other that is going from $\vartheta_b$ to $\vartheta_{\bar{a}}=\pi$ is

\begin{equation}\label{eq:lambda3}
    \lambda(b\bar{a})=e^{\rho_0}\sin\left(\frac{\pi-\vartheta_b}{2}\right)=e^{\rho_0}\cos\frac{\vartheta_b}{2}
\end{equation}

Notice that the metric in \eqref{eq:metricD} is invariant under the transformation $z\rightarrow ze^{i\gamma}$ which is a rotation of the disk by an angle $\gamma$. Let us assume that we are only interested in the lambda length of a given geodesic with half opening angle $\alpha$. Then we can represent it by any disk geodesic, which has the same $\alpha$ and regularization, and is rotationally equivalent to the original arc. This confirms the fact, that the lambda length is independent of the midpoint angle $\theta$. 

This gives us the opportunity to generalize the expressions above. To the points $a,b\in\partial BTZ$ (such that $-\infty<t_a<t_b<\infty$) we associate two centrally symmetric pairs of points on $\partial\mathbb{D}$ \cite{Fomin3}. Denote these by $a,b,\bar{a},\bar{b}\in\partial\mathbb{D}$. If $a,b\in\partial BTZ$ has got a hyperbolic angles $t_a<t_b$ then we can calculate the complex arguments $0\leq\vartheta_a<\vartheta_b<\pi$ of $a,b\in\partial\mathbb{D}$ by \eqref{eq:angles} and then for $\bar{a},\bar{b}\in\partial\mathbb{D}$, $\vartheta_{\bar{a}}=\vartheta_a+\pi<\vartheta_{\bar{b}}=\vartheta_b+\pi$.

At this point it is convenient to introduce a notation for different boundary representations. We denote BTZ boundary intervals by square brackets, and $\partial\mathbb{D}$ intervals by round brackets. For example, let $a$ and $b$ be two distinct points on $\partial BTZ$ (such that $t_a<t_b$) determining two intervals $[ab],[ba]\subset\partial BTZ$. Let $a,\bar{a},b,\bar{b}$ their centrally symmetric representatives on $\partial\mathbb{D}$ and the intervals between them are $(ab),(\bar{a}\bar{b}),(b\bar{a}),(\bar{b}a),(a\bar{a}),(b\bar{b}),(\bar{a}a),(\bar{b}b)\subset \partial\mathbb{D}$. Now we can assign to each centrally symmetric pair of intervals between the four points on $\partial\mathbb{D}$ exactly one physical interval between the two points on $\partial BTZ$, namely:

\begin{subequations}
\begin{align}
    \partial\mathbb{D}\supset(ab),(\bar{a}\bar{b})&\sim[ab]\subset\partial BTZ \label{eq:intervals1}\\
    \partial\mathbb{D}\supset(b\bar{a}),(a\bar{b})&\sim[ba]\subset\partial BTZ \label{eq:intervals2}\\
    \partial\mathbb{D}\supset(a\bar{a}),(\bar{a}a)&\sim[\partial BTZ] \label{eq:intervals3}\\
    \partial\mathbb{D}\supset(b\bar{b}),(\bar{b}b)&\sim[\partial BTZ] \label{eq:intervals4}
\end{align}    
\end{subequations}
Where $[\partial BTZ]$ denotes the whole BTZ boundary.

\begin{figure}[t]
    \centering
    \includegraphics[width=0.7\columnwidth]{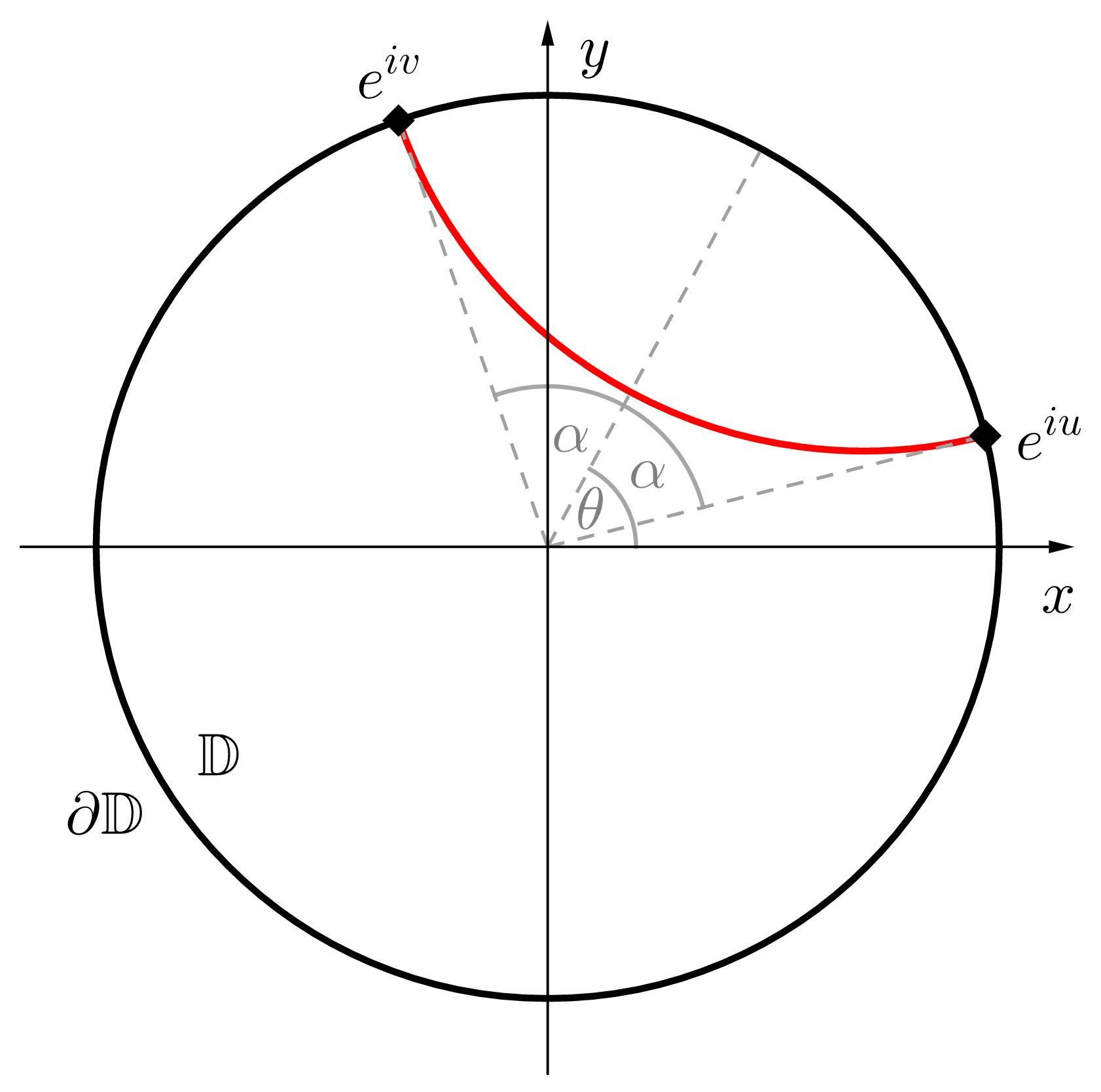}
    \caption{The Poincar\'e disk is the unit disk in the complex plane endowed with the \eqref{eq:metricD} metric. The red arc is a geodesic. These curves can be parametrized by $(\theta,\alpha)$ pairs. $\theta\in[0,2\pi]$ is the center, and $\alpha\in[0,\pi]$ is half the opening angle of the geodesic. Another useful parametrizations is given by the $(u,v)$ pair defined by Eq. \eqref{eq:uv}.}
    \label{fig:disk_geodesic}
\end{figure}

Notice that the lambda length of the geodesic between points $a,b \in\partial\mathbb{D}$ is the same as the lambda length between $\bar{a},\bar{b} \in\partial\mathbb{D}$, because they have the same opening angle (they differ only in a rotation around the origin). They also represent the same geodesic in the BTZ bulk. So the lambda length associated to an $[ab]$ geodesic in the BTZ bulk can be written in the following ways: $\lambda[ab]=\lambda(ab)=\lambda(\bar{a}\bar{b})$. Similarly $\lambda[ba]=\lambda(b\bar{a})=\lambda(\bar{b}a)$ and for loop geodesics $\lambda[\partial BTZ]=\lambda(a\bar{a})=\lambda(\bar{a}a)=\lambda(b\bar{b})=\lambda(\bar{b}b)$. The explicit expressions for these quantities can be given by equations \eqref{eq:lambda1}, \eqref{eq:lambda2} and \eqref{eq:lambda3}. If $\alpha=(\vartheta_b-\vartheta_a)/2$, then

\begin{subequations}
\begin{align}
    \lambda[ab]=\lambda(ab)=\lambda(\bar{a}\bar{b})&=e^{\rho_0}\sin\alpha\\
    \lambda[ba]=\lambda(b\bar{a})=\lambda(\bar{b}a)&=e^{\rho_0}\cos\alpha\\
    \lambda[\partial BTZ]=\lambda(a\bar{a})=\lambda(\bar{a}a)&=e^{\rho_0}\sin\frac{\pi}{2}\\
    \lambda[\partial BTZ]=\lambda(b\bar{b})=\lambda(\bar{b}b)&=e^{\rho_0}\sin\frac{\pi}{2}
\end{align}    
\end{subequations}
These lambda lengths fully describe quantitatively the geodesic structure of a doubly marked BTZ black hole in the HTL (see \hyperref[fig:geod_representations]{FIG. 3c.}).

We make one last remark to this section, namely that the usual expression for the geodesic length calculated by the line element \eqref{eq:static_btz} is of the form
\begin{equation}
\label{eq:btzlength}
\ell[ab]=2R\log \left(\frac{2r_0}{r_+} \sinh \frac{t_b-t_a}{2}\right)
\end{equation}
Where $r_0$ is some regularization factor taken to be large. By virtue of the \eqref{eq:kell} 
relation $r_0=r_+\cosh\varrho_0$ for $\varrho_0$ large one has $\frac{2r_0}{r^+}=e^{\varrho_0}$. 
Then this formula can be connected to \eqref{eq:disk_length} via \eqref{eq:angles}, because one can rewrite the hyperbolic factor in the argument of the logarithm in the following way

\begin{equation}
\label{eq:kapocs}
\begin{aligned}
    \sinh\frac{t_b-t_a}{2}&=\frac{1}{2}e^{-(t_b+t_a)/2}\left(e^{t_b}-e^{t_a}\right)=\\
    &=\frac{\cot{(\vartheta_b/2)}-\cot{(\vartheta_a/2)}}{2\sqrt{\cot{(\vartheta_b/2)}\cot{(\vartheta_a/2)}}}=\\
    &=\frac{1}{\sqrt{\sin{\vartheta_b}\sin{\vartheta_a}}}\sin{\frac{\vartheta_b-\vartheta_a}{2}}
\end{aligned}
\end{equation}
or alternatively

\begin{equation}
\label{eq:kapocs2} 
\sin{\frac{\vartheta_b-\vartheta_a}{2}}=
\frac{1}{\sqrt{\cosh{t_b}\cosh{t_a}}}
\sinh\frac{t_b-t_a}{2}
\end{equation}

Now one can compare \eqref{eq:btzlength} and  \eqref{eq:disk_length}. The difference between the the two formulas comes from the fact, that one can use uniform regularization either in the BTZ representation
or in the disk representation but not both. Indeed \eqref{eq:kapocs} and \eqref{eq:kapocs2} show that when going from one representation to the other the regularization will be either $t$ or $\vartheta$ dependent. It can be shown that this corresponds to regularizations via horocycles with different diameters at the points $a$ and $b$.
Notice, however that there exist gauge invariant (regularization independent) combinations of lambda lengths like cross ratios where these subtleties are immaterial. 
These combinations give rise to physical quantities of great importance.
Next we turn to such quantities. 

\section{Entropies and conditional informations}

According to the Ryu-Takayanagi proposal the entanglement entropy of a given domain $A$ is proportional to the area of the minimal surface that has the same boundary as $A$. In the static case this means that

\begin{equation}\label{eq:entropy_ell}
    S(A)=\frac{\ell(A)}{4G}
\end{equation}
where $A$ is a boundary interval and $\ell(A)$ is the regularized length of the geodesic homologous to $A$.
With the help of lambda lengths defined in \eqref{eq:lambda_D} we can reformulate this expression
as follows

\begin{equation}\label{eq:entropy_lambda}
    S(A)=\frac{c}{3}\log\lambda(A)
\end{equation}
Where $c$ is the central charge of the boundary theory expressed by the Brown-Henneaux relation \cite{BH} $c=\frac{3R}{2G}$.

In the BTZ black hole case studied here there is an important caveat. The expressions as given by \eqref{eq:entropy_ell} and
\eqref{eq:entropy_lambda} are only valid in the high temperature limit i.e. when $M\gg 1$. Indeed, for a finite black hole mass there exists a critical size for the region past which there is a new family of disconnected geodesics that have smaller length then the connected homologous ones\cite{HT,EP}. 
In this case the prescription \eqref{eq:entropy_lambda} is ambiguous. However, in the high temperature limit (HTL) the critical size of the region (the entanglement plateaux scale\cite{EP}) measured by half the opening angle of that region is $\pi$ hence we did not need to worry about this subtlety. Moreover, in this case the Ryu-Takayanagi geodesics reach all the way to the horizon (the smallest entanglement shadow occurs in the HTL\cite{Zukowski}).
Hence in the following we restrict attention to the macroscopic BTZ black hole showing up in this HTL.

Now assume that we choose two intersecting regions labeled by $E,F$ on the BTZ boundary. These two regions give rise to the following CFT subsystems denoted by $A=E\setminus F$, $B=E\cap F$, $C=F\setminus E$ and $D=E\cup F$. According to strong subadditivity, for the corresponding von-Neumann entropies one has\cite{NC}

\begin{equation}\label{eq:strong_sub}
    S(AB)+S(BC)-S(B)-S(D)\geq0
\end{equation}
If we introduce the conditional entropy

\begin{equation}
    S(A|B)\equiv S(AB)-S(B)
\end{equation}
Mutual information

\begin{equation}
    I(A,B)\equiv S(A)-S(A|B)
\end{equation}
And the conditional mutual information

\begin{equation}
\begin{aligned}
    I(A,C|B)&\equiv S(A|B)-S(A|BC)=\\
    &=I(A,BC)-I(A,B)
\end{aligned}
\end{equation}
Then we can rewrite \eqref{eq:strong_sub} as

\begin{equation}
    I(A,C|B)=S(A|B)-S(A|BC)\geq 0
\end{equation}
Hence strong subadditivity indicates that conditioning on a larger subsystem can only reduce the uncertainty about a system.

Due to the Ryu-Takayanagi conjecture, we can also express these entropic quantities in terms of lambda lengths. The conditional entropy is

\begin{equation}
    S(A|B)=\frac{c}{3}\log\frac{\lambda(E)}{\lambda(B)}
\end{equation}
The mutual information
\begin{equation}
    I(A,B)=\frac{c}{3}\log\frac{\lambda(A)\lambda(B)}{\lambda(E)}
\end{equation}
And the conditional mutual information is

\begin{equation}\label{eq:cond1}
    I(A,C|B)=\frac{c}{3}\log\frac{\lambda(E)\lambda(F)}{\lambda(B)\lambda(D)}
\end{equation}

Now let us turn to the Poincar\'e disk representation. Label the BTZ boundary points by $a,b,c,d\in\partial BTZ$. Then the corresponding regions are $A=[ab]$, $B=[bc]$ and $C=[cd]$. Each point on $\partial BTZ$ gives rise to a centrally symmetric pair of points on $\partial\mathbb{D}$. Let us denote them by $a,\bar{a},b,\bar{b},c,\bar{c},d,\bar{d}\in\mathbb{D}$. Next as in \eqref{eq:intervals1}, \eqref{eq:intervals2}, \eqref{eq:intervals3} and \eqref{eq:intervals4}
we associate to BTZ boundary regions disk boundary ones.
Since for every $\partial\mathbb{D}$ region there is a homologous $\mathbb{D}$ geodesic then in the disk representation one can explicitly write down the expressions for the lambda lengths and von-Neumann entropies. In HTL the exact result for the entropy of subsystem $A$ is

\begin{equation}\label{eq:entropy_disk}
\begin{aligned}
    S[A]&=S(A)=\\
    &=\frac{c}{3}\log e^{\rho_0}\sin\left(\alpha\right)=\\
    &=\frac{c}{3}\log e^{\rho_0}\sin\left(\frac{\vartheta_b-\vartheta_a}{2}\right)
\end{aligned}
\end{equation}
Where $\vartheta_a,\vartheta_b$ are the complex arguments corresponding to endpoints of the regions $(ab)\subset\partial\mathbb{D}$. We can say that $S(A)=S(ab)=S(\bar{a}\bar{b})$, because both regions represent the same subsystem on the BTZ boundary. Finally we record that in this doubly marked disk representation the entanglement entropy of the entire boundary is $S[\partial BTZ]=S(a\bar{a})=S(b\bar{b})$.

We can calculate similarly the expressions for the other entropic quantities. 
For example $S(A|B)$ can be calculated as

\begin{equation}
    S[A|B]=S(A|B)=\frac{c}{3}\log\frac{\sin\left(\frac{\vartheta_c-\vartheta_a}{2}\right)}{\sin\left(\frac{\vartheta_c-\vartheta_b}{2}\right)}
\end{equation}

An explicit formula, to be used later, for the mutual information is given by
\begin{equation}
    I[A,B]=I(A,B)=\frac{c}{3}\log e^{\rho_0}\frac{\sin\left(\frac{\vartheta_b-\vartheta_a}{2}\right)\sin\left(\frac{\vartheta_c-\vartheta_b}{2}\right)}{\sin\left(\frac{\vartheta_c-\vartheta_a}{2}\right)}
\end{equation}

and for the conditional mutual information of \eqref{eq:cond1} 

\begin{equation}\label{eq:mutual}
\begin{aligned}
     I[A,C|B]&=I(A,C|B)=\\
     &=\frac{c}{3}\log\frac{\sin\left(\frac{\vartheta_c-\vartheta_a}{2}\right)\sin\left(\frac{\vartheta_d-\vartheta_b}{2}\right)}
    {\sin\left(\frac{\vartheta_c-\vartheta_b}{2}\right)\sin\left(\frac{\vartheta_d-\vartheta_a}{2}\right)}
\end{aligned}
\end{equation}

Notice that the conditional mutual information is finite, and independent of the choice of regularization, while the the entanglement entropy and mutual information are divergent and they depend on the regularization. 

Having introduced our basic quantities of quantum information in the disk representation, let us elucidate the meaning some of these in the BTZ representation.
First of all in the usual BTZ representation we have the following formula for the entanglement entropy\cite{Calabrese}
\begin{equation}
\label{eq:usual}
S[A]=\frac{c}{3}\log\left(\frac{\beta}{\pi \epsilon}\sinh\left(\frac{\pi\ell}{\beta}\right)\right)
\end{equation}
where $\ell =\Delta\varphi r_0$, $L=2\pi r_0$  and $\sqrt{M}=\frac{r_+}{R}=\frac{L}{\beta}$.
One can also write 
\begin{equation}
\label{eq:reszformula}
\frac{\pi\ell}{\beta}=\frac{1}{2}\sqrt{M}\Delta\varphi=\frac{1}{2}\Delta t
\end{equation}
where we have used \eqref{eq:kell} and
$\Delta t=t_a-t_b$.
Using now $\frac{2r_0}{r_+}\simeq\frac{\beta}{\pi\epsilon}$ and formula 
\eqref{eq:kapocs} one can see that expressions \eqref{eq:usual} and \eqref{eq:entropy_disk} can be converted to each other by applying either  $t$ or $\vartheta$ dependent cutoffs.
Of course when calculating conditional mutual informations such cutoffs are immaterial, since these quantities are gauge (regularization) invariant ones.
Indeed, an alternative formula for \eqref{eq:mutual} is then given by

\begin{equation}\label{eq:mutual2}
     I(A,C|B)=\frac{c}{3}\log\frac{\sinh\left(\sqrt{M}\frac{\varphi_c-\varphi_a}{2}\right)\sinh\left(\sqrt{M}\frac{\varphi_d-\varphi_b}{2}\right)}
    {\sinh\left(\sqrt{M}\frac{\varphi_c-\varphi_b}{2}\right)\sinh\left(\sqrt{M}\frac{\varphi_d-\varphi_a}{2}\right)}
\end{equation}

Let us consider now in the BTZ picture the special case when in the disc picture we have $c=\overline{a}$ and $d=\overline{b}$.
In this case $\varphi_c-\varphi_a=\varphi_d-\varphi_b=2\pi$, and $\Delta\varphi\equiv\varphi_c-\varphi_b=\varphi_d-\varphi_a$ (see FIG 3c).
Then in this special case to be used later we have
\begin{equation}
\label{eq:later}
\begin{aligned}
I(ab,\overline{a}\overline{b}|b\overline{a})&=I[A,C|B]=\\
&=\frac{2c}{3}\left[\log\left(\sinh\pi\sqrt{M}\right)-\log\left(\sinh\frac{\Delta\varphi}{2}\sqrt{M}\right)\right]
\\&=
\frac{c}{3}(2\pi-\Delta\varphi)\sqrt{M}+\dots\\& =2\frac{(2\pi-\Delta\varphi)r_+}{4G}+\dots
\end{aligned}
\end{equation}
where the dots refer to terms vanishing in the high temperture limit, $\sqrt{M}=\frac{r_+}{R}\gg 1$ and  $c=\frac{3R}{2G}$  due to the Brown-Henneaux relation.

\section{The cluster algebraic structure of BTZ triangulations}

\begin{figure*}[t]
    \hspace{1cm}
    \subfloat[]{
        \includegraphics[width=0.41\columnwidth]{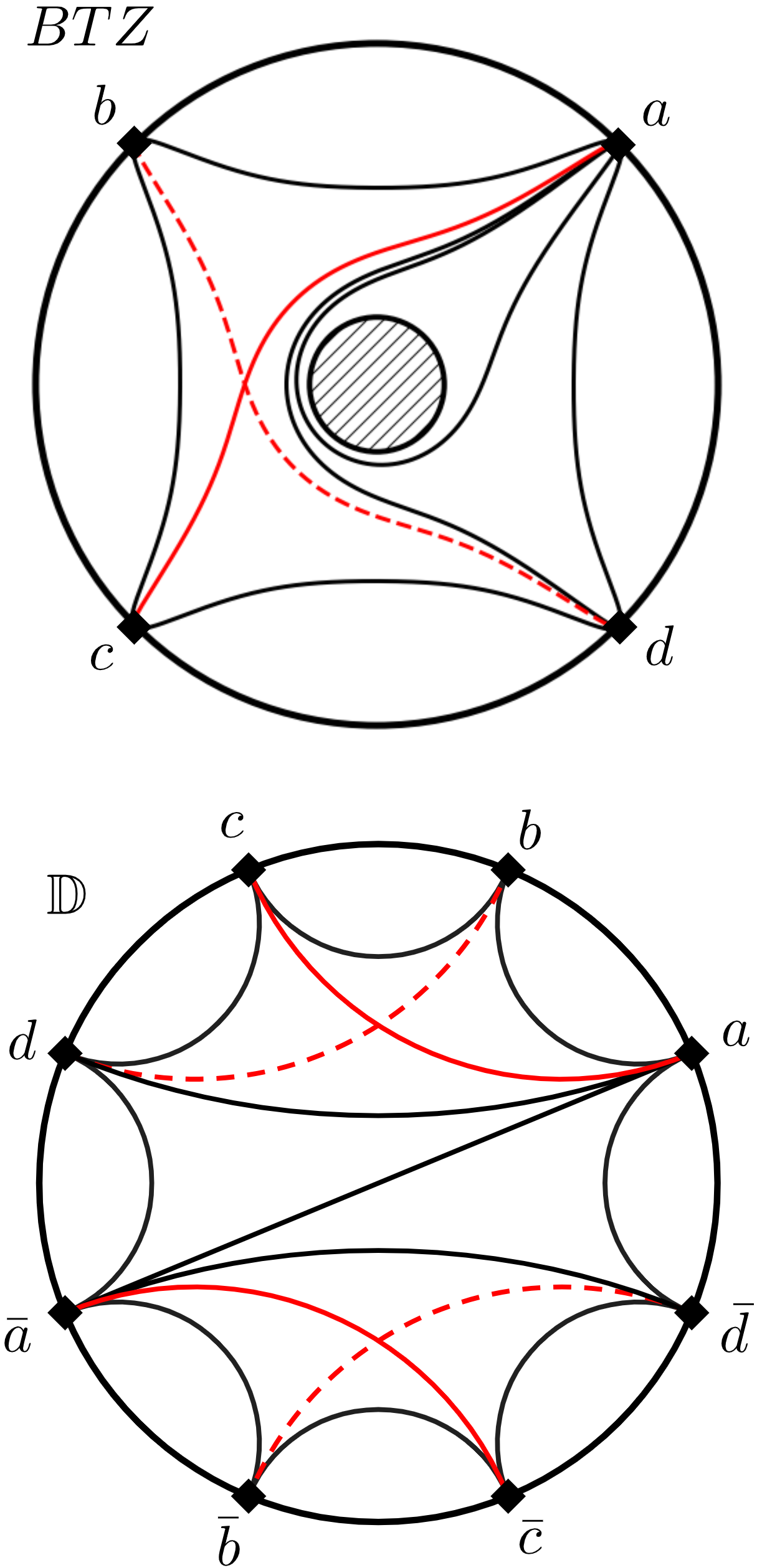}}
    \hfill
    \subfloat[]{
        \includegraphics[width=0.4\columnwidth]{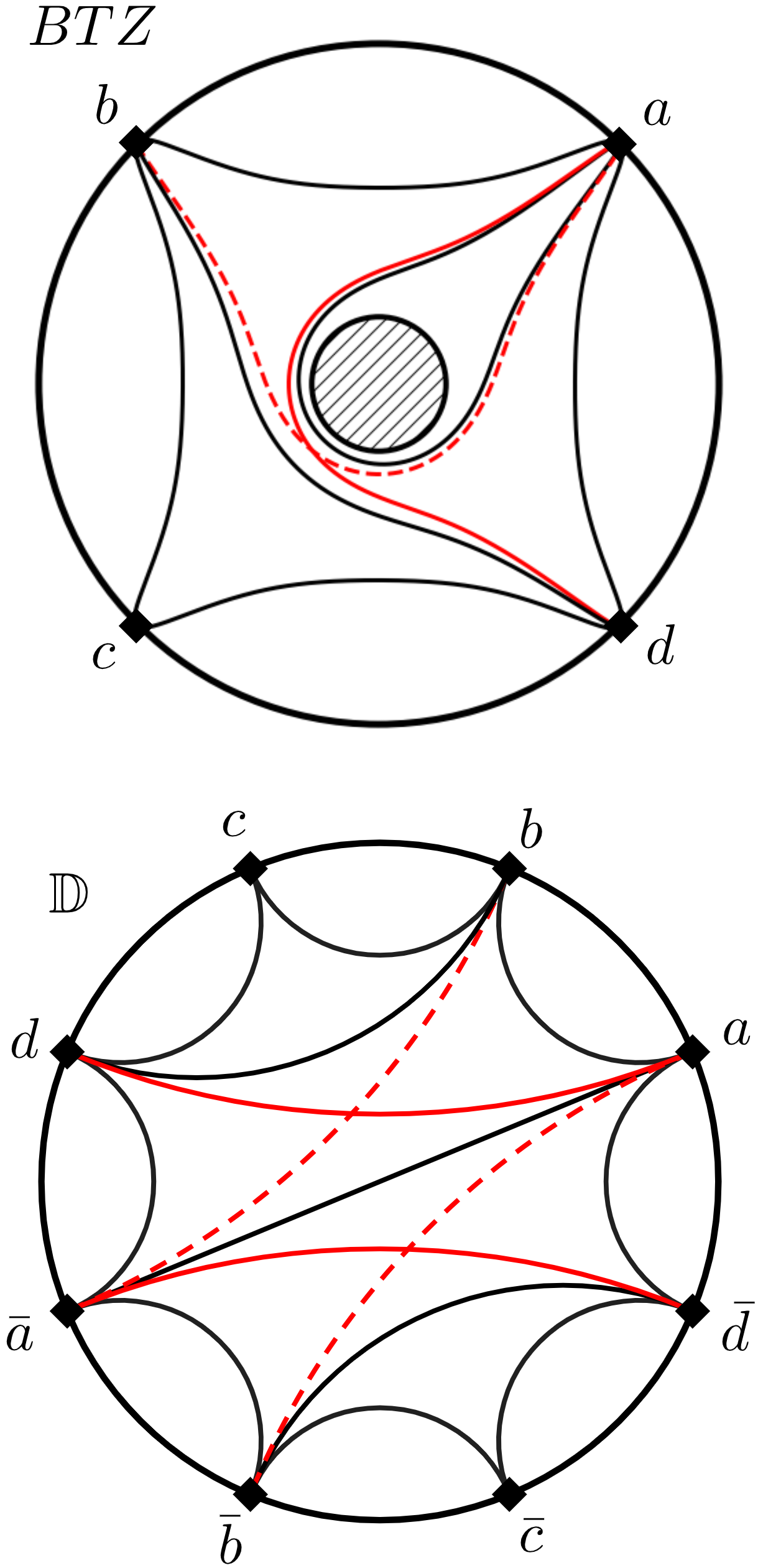}}
    \hfill
    \subfloat[]{
        \includegraphics[width=0.42\columnwidth]{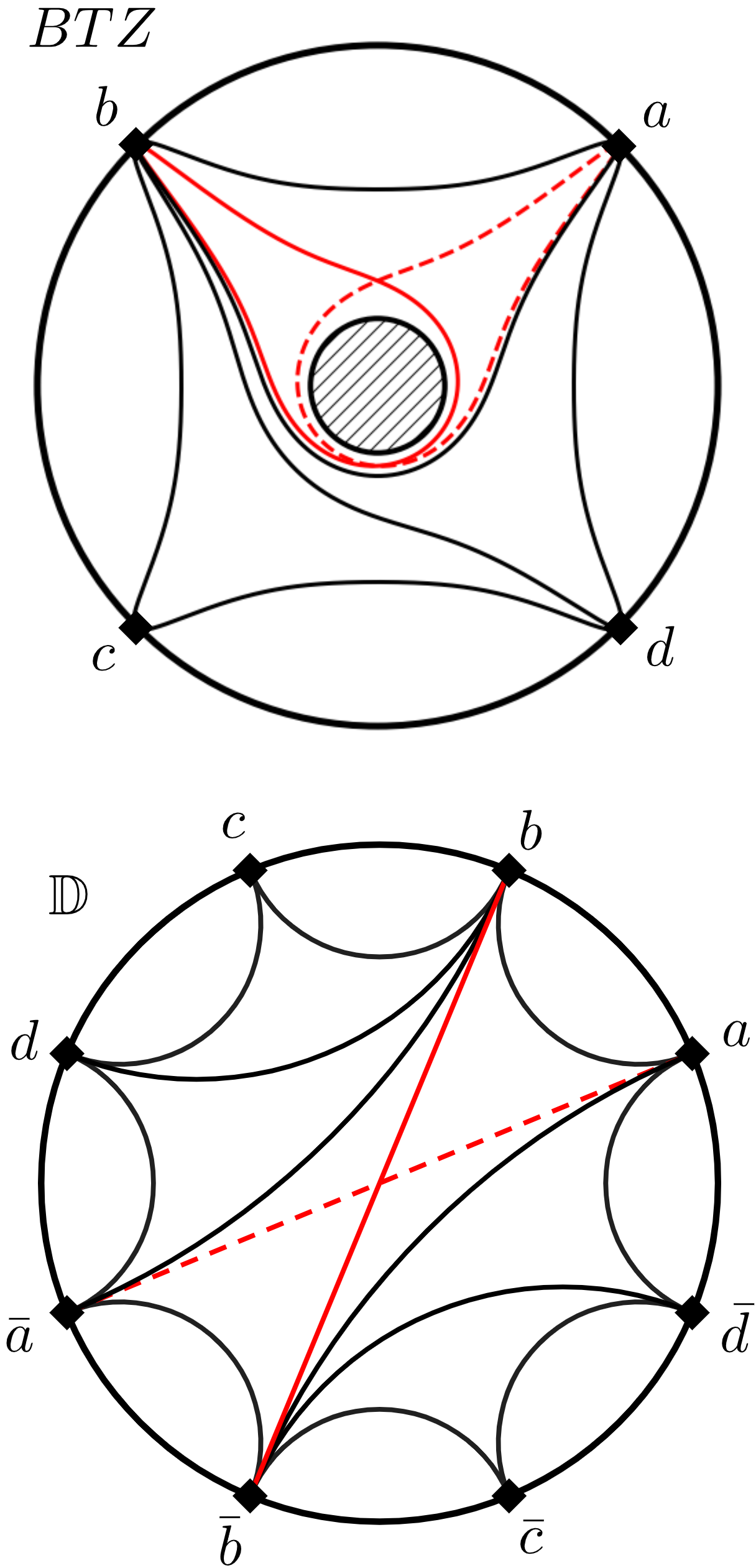}}
    \hspace{1cm}
    \caption{Triangulations of a BTZ black hole with $N=4$ marked points on the boundary (top figures), and their $2N=8$-gon disk representations (bottom figures). Solid black and red curves are the edges and diagonals of the polygons. Each centrally symmetric pair of non-diametric diagonals (or one simple diameter) in the disk corresponds to a diagonal of the BTZ quadrangle. In each picture we illustrated a flip from the solid red curves to the dashed red ones. In (a) there is a flip in an ordinary quadrangle. The exchange relations for this type of flip are given by \eqref{eq:flip1} and \eqref{eq:flip1_BTZ} with labels $i=a$, $j=b$, $k=c$ and $l=d$. In (b) a diagonal of a folded quadrangle is flipped. The exchange relations are \eqref{eq:flip4} and \eqref{eq:flip2_BTZ} with labels $i=a$, $j=b$ and $k=d$. Finally (c) shows a loop flip. In this case the exchange relations are given by \eqref{eq:flip6} and \eqref{eq:flip3_BTZ} with labels $i=a$ and $j=b$.}
    \label{fig:flips}
\end{figure*}

Now we are in a position to present the main result of this paper.
We show that geodesic triangulations of the BTZ black hole in the HTL, with $N$ marked points on the boundary, exhibit a particular structure. This structure manifests itself in an exchange pattern called $C_{N-1}$ well known from the literature on cluster algebras. Since via the Ryu-Takayanagi formula regularized geodesic lengths are directly related to entanglement entropies, this $C_{N-1}$ structure also provides an algebraic characterization of the entanglement patterns of the boundary thermal state dual to the BTZ geometry.

Before presenting a detailed elaboration of this result, let us recall some basic definitions\cite{Williams}. Fix $N$ points on the $r\rightarrow\infty$ boundary. A triangulation of the bulk is a maximal set of distinct, pairwise non-intersecting geodesics. We assume that these geodesics do not intersect themselves in the bulk. They split the bulk geometry into domains, which we call triangles. Now assume that we delete an arc from a given triangulation and add another one instead such that we get a new triangulation. We call this transformation a flip. Let the number of compatible arcs in the triangulations be $n$. Then we call the $n$-regular graph with the triangulations as its vertices an exchange graph, if its vertices are connected by the flips between the corresponding triangulations.

Now we turn to the static, macroscopic BTZ black hole. Its outer horizon region is not simply connected. Recall from \hyperref[sec:3]{Section III.}, as a result of this in the BTZ geometry we find different types of geodesics. There are geodesic arcs that wind around the hole and having coinciding boundary endpoints. Such arcs are called loops. On the other hand between two different marked boundary points there can be two ordinary arcs: one of them located on one side of the hole and another one on the other. 

A triangulation of the BTZ black hole with $N$ marked point consists of $N-2$ ordinary geodesics and a loop. If we triangulate the bulk, then it is built from the following domains. First of all there are ordinary triangles whose edges are sides or diagonals of the polygon. Then we also have folded triangles with one of its edges being the loop (so it has got two vertices). Finally we have a non-simply connected domain containing the hole, with the domain boundary formed by the loop and the horizon. In this case there are two types of flips. First we can flip a diagonal in a quadrangle. This quadrangle can also be folded if one of its sides is the loop. Second we can also flip the loop in a digon to get an other loop. The three examples of such flips for $N=4$ are illustrated in \hyperref[fig:flips]{FIG. 4.}

\begin{figure*}[t]
    \hspace{1cm}
    \subfloat[]{
        \includegraphics[width=0.7\columnwidth]{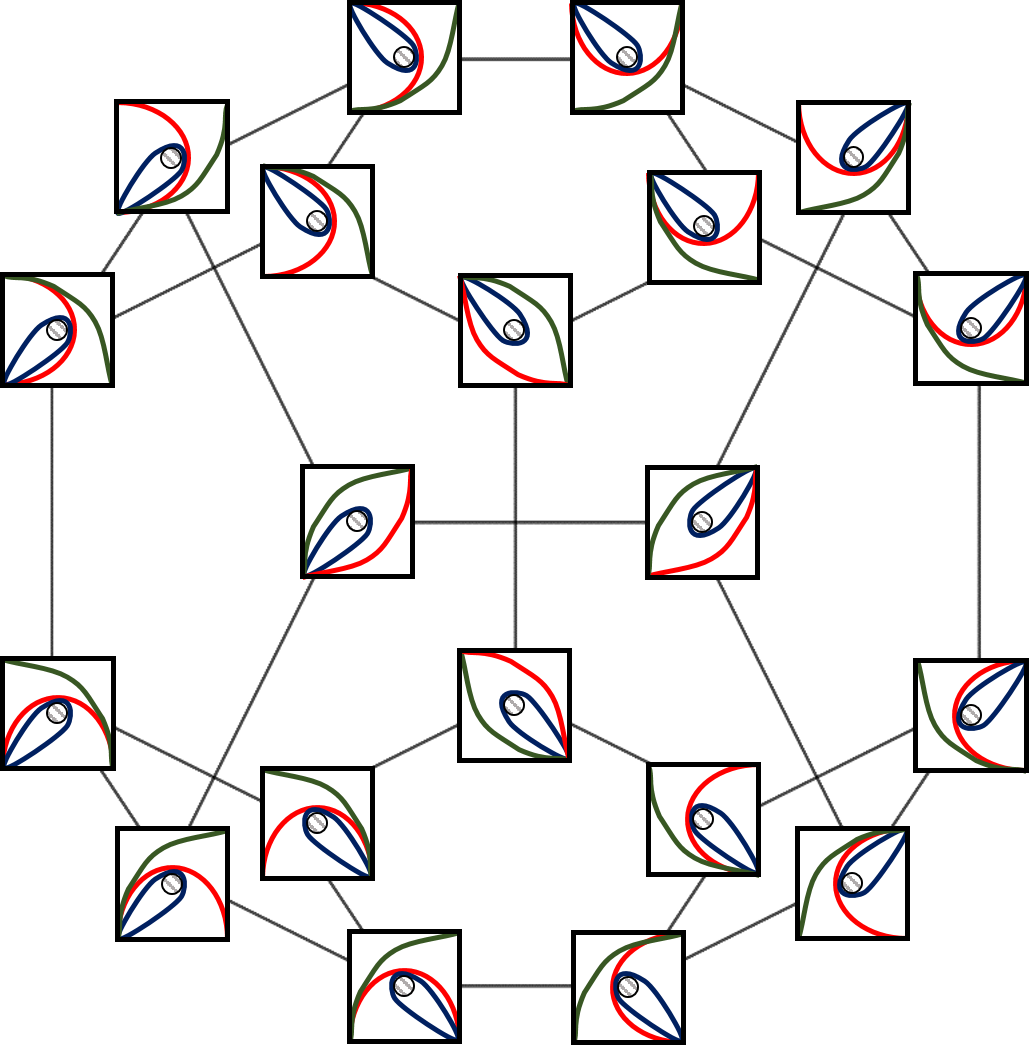}}
    \hfill
    \subfloat[]{
        \includegraphics[width=0.7\columnwidth]{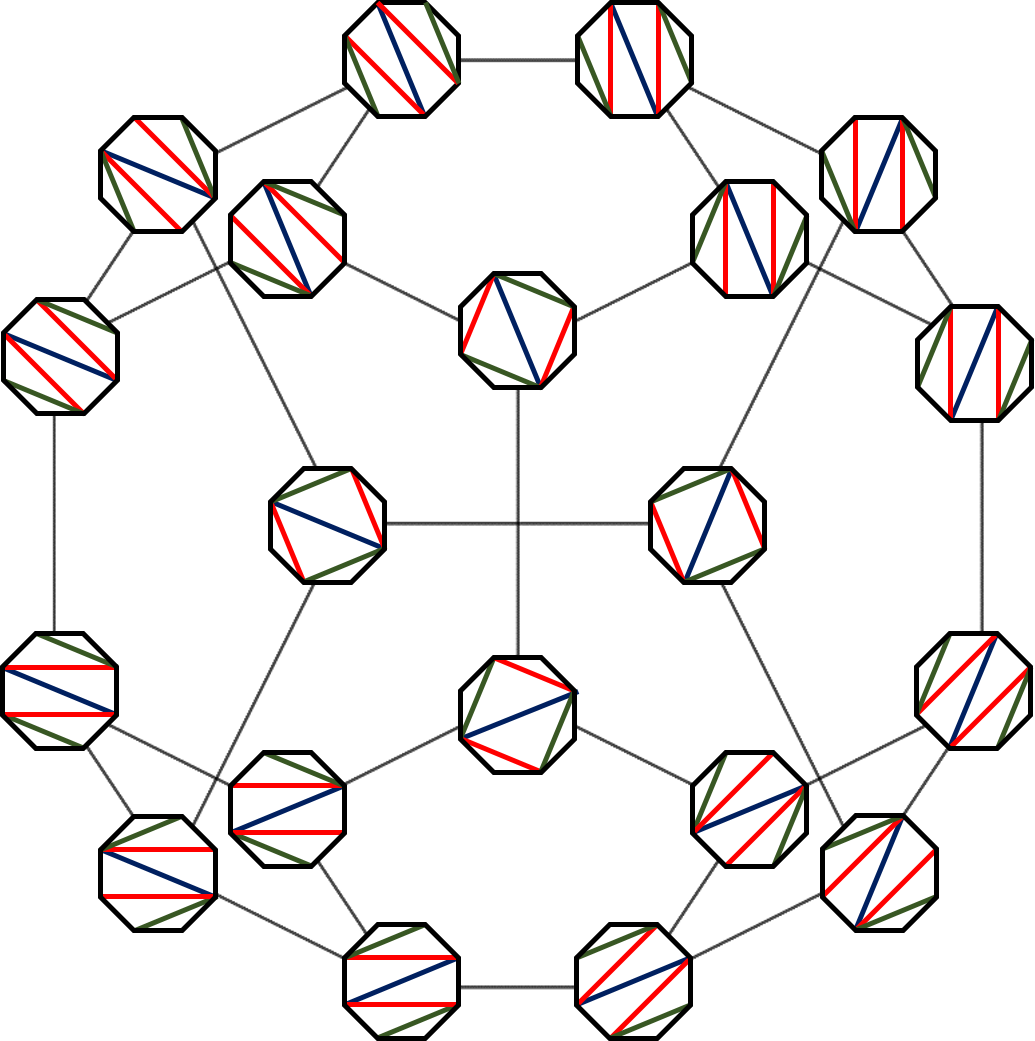}}
    \hspace{1cm}
    \caption{The cyclohedron $\mathcal{C}_3$ is the exchange graph of a $C_{3}$ cluster algebra. In (a) the vertices are the triangulations of a quadrangle with a hole inside. This illustrates the triangulations of the BTZ black hole with $N=4$ marked points on the boundary. In (b) the vertices are the centrally symmetric octagon equivalents of the triangulations in (a). This meets with the $2N=8$ disk representation of the BTZ quadrangle. Each pairs of non-diametric diagonal with one color represent one single diagonal in the quadrangle picture with the same color.}
    \label{fig:cyclo}
\end{figure*}

We can construct the exchange graph of the triangulation if we notice that there is a one-to-one correspondence between the triangulations of an $N$-gon with a hole in its center, and the centrally-symmetric triangulations of an ordinary $2N$-gon. Let us denote the $N$ marked points on the upper half $\partial \mathbb{D}$ by $a,b,\dotsc$. We can construct the $2N$-gon by adjoining an antipodal copy of each vertex of the $N$-gon denoted by $\bar{a},\bar{b},\dotsc$. Then every ordinary arc of the $N$-gon will correspond to a centrally symmetric pair of diagonals $(i\bar{j})$ and $(\bar{i}j)$ ($i$ and $j$ labels different vertices) and the loop corresponds to a diameter $(i\bar{i})$ of the $2N$-gon. Then an ordinary flip of the $N$-gon is a centrally-symmetric flip of two diagonals, and the exchange of the loop is the flip of the diameter in the picture of the $2N$-gon. With this construction the resulting exchange graph is a $\mathcal{C}_{N-1}$ cyclohedron \cite{BT,HL,Devadoss}. For $N=4$ see \hyperref[fig:cyclo]{FIG. 5}.
Then one can conclude that the exchange pattern of the static, macroscopic BTZ black hole is a cyclohedron.

There is a natural correspondence between triangulations of bordered surfaces and cluster algebras\cite{Williams}. In the case of the entanglement patterns of the vacuum state dual to pure $AdS_3$ we already know that the associahedron is not only a special case for an exchange graph of such triangulations, but it is generated by a seed pattern of an $A_{N-3}$ cluster algebra\cite{LB}. Hence one can pose the question: Is there also a cluster algebra behind the thermal state dual to the BTZ black hole encoding entanglement patterns via our cyclohedron?

It is known from the theory of cluster algebras that the cyclohedron is the seed pattern for $C_n$ type cluster algebras. Hence one can conjecture that a $C_n$ type cluster algebra governs the algebraic properties of the geodesics of the static, HTL BTZ black hole. Moreover, since this bulk black hole is dual to a boundary thermal state, one can argue via the Ryu-Takayanagi correspondence that the same algebraic structure also governs the entanglement patterns of this quantum state. 

In order to elaborate on this conjecture first we mark $N=4$ points (label them by $a,b,c,d$) on the boundary of 
the Poincar\'e disk. We can think of them forming a quadrilateral with sides $(ab)$,$(bc)$,$(cd)$ and $(ad)$. Then the following relation holds between the corresponding geodesic lambda lengths \eqref{eq:lambda_D}:

\begin{equation}\label{eq:Ptolemy}
    \lambda(ac)\lambda(bd)=\lambda(ab)\lambda(cd)+\lambda(ad)\lambda(bc)
\end{equation}
This is the Ptolemy relation\cite{Penner} which is true for any arbitrary geodesic quadrangle on the whole Poincar\'e disk.

Now we cut the disk in half and make the $\vartheta=0\sim\vartheta=\pi$ identification to get the representation of our BTZ black hole. If we mark $N$ points on the BTZ boundary, then in this disk representation each point will be located on the upper semicircle of the disk.

We already know that to each $a,b,c,\dotsc\in\partial BTZ$ (in cyclic order) vertex of a BTZ $N$-gon there corresponds a centrally symmetric pair of vertices in the corresponding $2N$-gon picture. Denote this new set of points by $a,b,c,\dotsc,\bar{a},\bar{b},\bar{c},\dotsc\in\partial\mathbb{D}$. As in \eqref{eq:intervals1}, \eqref{eq:intervals2}, \eqref{eq:intervals3} and \eqref{eq:intervals4}, for every ordinary geodesic $[ij]\subset\partial BTZ$ corresponds a centrally symmetric pair of geodesics $(ij),(\bar{i}\bar{j})\subset\partial\mathbb{D}$ (or $(i\bar{j}),(j\bar{i})\subset\partial\mathbb{D}$), and for every loop like geodesic $[ii]\subset\partial BTZ$ corresponds a diameter $(i\bar{i})\subset\partial\mathbb{D}$ of the $2N$-gon.

Every flip in a triangulation of the $N$-gon corresponds to a centrally symmetric (or a diametrical) flip of the $2N$-gon. Choose four different $i,j,k,l\in\partial\mathbb{D}$ points (in cyclic order) from $a,b,c,\dots\in\partial\mathbb{D}$. With the Ptolemy relation in \eqref{eq:Ptolemy} and the lambda lengths of the orbits we can give six exchange relations to different types of flips 
\begin{subequations}
\begin{align}
    &\lambda(i k) \lambda(j l)=\lambda(i j) \lambda(k l)+\lambda(i l) \lambda(j k) \label{eq:flip1}\\
    &\lambda(k \bar{i}) \lambda(j l)=\lambda(j \bar{i}) \lambda(k l)+\lambda(l \bar{i}) \lambda(j k)\label{eq:flip2}\\
    &\lambda(k \bar{i}) \lambda(l \bar{j})=\lambda(\bar{i} \bar{j}) \lambda(k l)+\lambda(l \bar{i}) \lambda(k \bar{j})\label{eq:flip3}\\
    &\lambda(i k) \lambda(j \bar{i})=\lambda(i j) \lambda(k \bar{i})+\lambda(i \bar{i}) \lambda(j k)\label{eq:flip4}\\
    &\lambda(j \bar{i}) \lambda(k \bar{j})=\lambda(\bar{i} \bar{j}) \lambda(j k)+\lambda(j \bar{j}) \lambda(k\bar{i})\label{eq:flip5}\\
    &\lambda(i \bar{i}) \lambda(j \bar{j})=\lambda(i j)\lambda(\bar{i} \bar{j})+\lambda(j \bar{i})\lambda(\bar{j} i)\label{eq:flip6}
\end{align}
\end{subequations}
Now one can compare this set of relation with the ($r=2$) set of equations (12.10)-(12.15) of Ref.\cite{Fomin3}.
After taking into account \eqref{eq:intervals1} and  \eqref{eq:intervals2} one can realize that these
are the exchange relations of a $C_{N-1}$ cluster algebra\cite{Fomin3}. This means that for the triangulations of the macroscopic BTZ black hole with $N$ marked points on its boundary, the geodesic lambda lengths generate a $C_{N-1}$ cluster algebra.

Using \eqref{eq:intervals2}, \eqref{eq:intervals3} and \eqref{eq:intervals4}, we can express the lambda lengths in terms of the BTZ boundary intervals as well. One can see that the first three relations correspond to the same type of flip (flip in an ordinary quadrangle). Similarly \eqref{eq:flip4} and \eqref{eq:flip5} are the algebraic relations for flips in folded quadrangles. Finally the last one corresponds to loop flips. Hence in terms of BTZ intervals we have got only the following three different exchange relations

\begin{subequations}
\begin{align}
    &\lambda[i k] \lambda[j l]=\lambda[i j] \lambda[k l]+\lambda[i l] \lambda[j k] \label{eq:flip1_BTZ}\\
    &\lambda[i k] \lambda[j i]=\lambda[i j] \lambda[k i]+\lambda[\partial BTZ] \lambda[j k]\label{eq:flip2_BTZ}\\
    &\lambda[\partial BTZ]^2=\lambda[i j]^2+\lambda[j i]^2\label{eq:flip3_BTZ}
\end{align}
\end{subequations}
where now $i,j,k,l\in\partial BTZ$ label $\partial BTZ$ points in cyclic order.

\begin{figure}[t]
    \centering
    \includegraphics[width=\columnwidth]{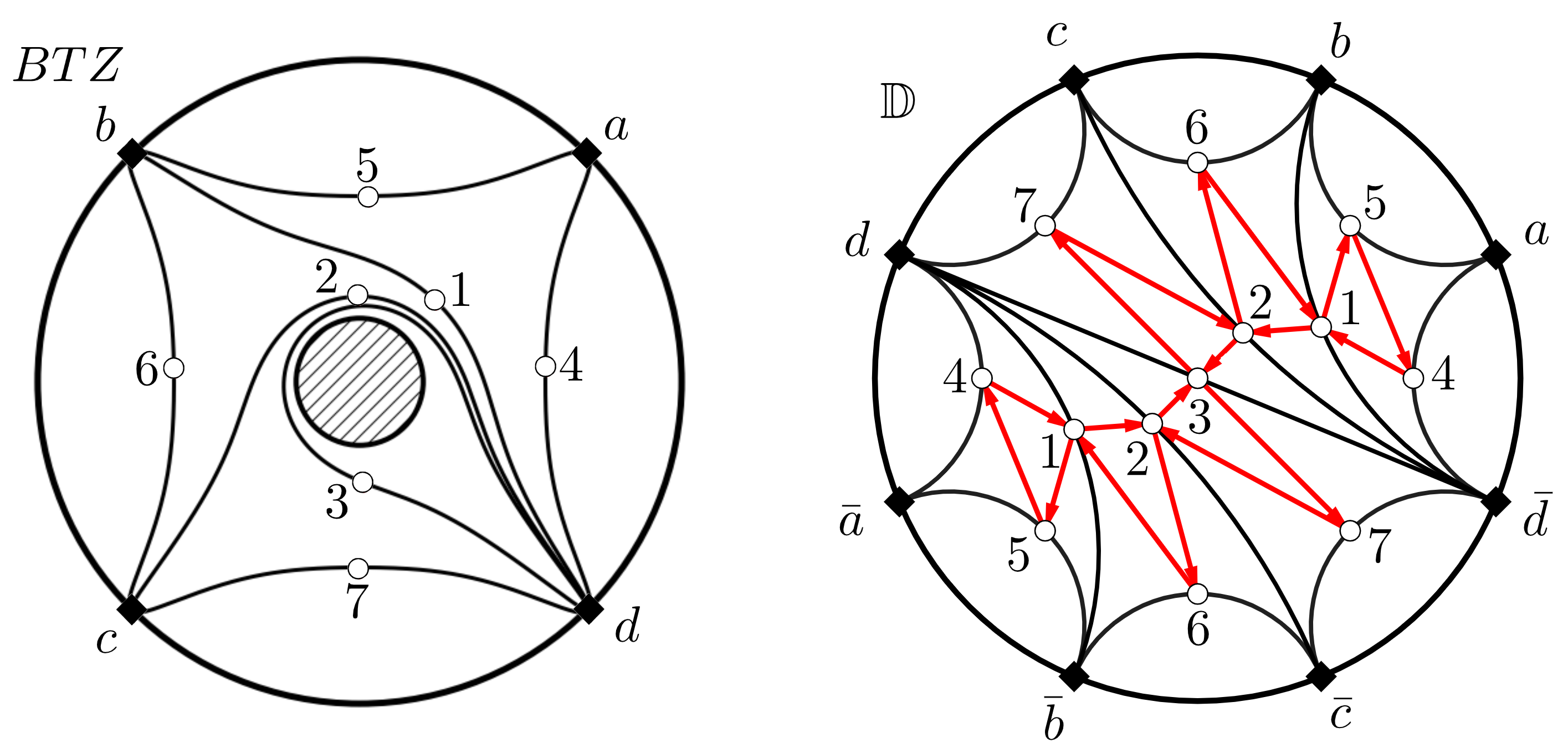}
    \caption{Graph representation of a BTZ quadrangle triangulation. On the left hand side the triangulation is in the BTZ picture. The diagonals are labeled by $1,\dots,N-1$ while the edges by $N,\dots,2N-1$. On the right hand side one can find the corresponding $2N$-gon representation in $\mathbb{D}$. Here the edges and diagonals are labeled according to \eqref{eq:intervals1},\eqref{eq:intervals2},\eqref{eq:intervals3} and \eqref{eq:intervals4}. These give rise to a directed graph, based on inscribed triangles, whose vertices correspond to geodesics. The edges of the graph (red arrows) belong to clockwise oriented inscribed triangles.}
    \label{fig:quiver}
\end{figure}

We can rewrite these relations in a more general form by encoding the geometry of a triangulations in an incidence matrix. In order to do this we choose an arbitrary BTZ triangulation and put a mark at the midpoint of its $[ij]\subset\partial BTZ$ diagonals and edges. We label the diagonals of the triangulation by $1,\dots,N-1$, and its edges by $N,\dots,2N-1$. In the disk $2N$-gon representation label $(ij)$ and $(j\bar{i})$ (or $(ij)$ and $(j\bar{i})$ if $j<i$) with the same number as $[ij]$. Now within each $\mathbb{D}$ triangle of the centrally symmetric $2N$ triangulation connect the markers of its sides. In this way we have obtained new inscribed triangles. Orient these new triangles clockwise. In this way we have obtained a directed graph whose vertices represent diagonals and edges of a given centrally symmetric disk triangulation (see \hyperref[fig:quiver]{FIG. 6.}).

Now we construct a $(2N-1)\times(2N-1)$ matrix $B$  with matrix elements $b_{ij}$ in the following way. Assume that $i$ and $j$ label different edges in the disk triangulation. Then $b_{ij}=k>0$ if there are $k$ arrows in the graph going from a chosen vertex labeled by $i$ to $k$ different vertices labeled by $j$, and $b_{ij}=-k<0$ if there are $k$ edges going from $k$ different vertices labeled by $j$ to a chosen edge labeled by $i$. For example one can check that in FIG. 6. we have $b_{15}=1$, $b_{32}=-2$ and $b_{37}=2$. We see that 
via keeping track of the mutual positions of the edges and diagonals
the matrix $B$ fully characterizes a BTZ triangulation.

Assume that we are flipping the $j$'th ($j\in{1,\dots,N-1}$) diagonal. We denote its lambda length before the flip by $\lambda[j]$, and after the flip by $\lambda'[j]$. Then 
 for an arbitrary triangulation
we can rewrite equations \eqref{eq:flip1_BTZ},\eqref{eq:flip2_BTZ} and \eqref{eq:flip3_BTZ} in the compact form

\begin{equation}
    \lambda[j]\lambda'[j]=\prod_{1\leq i\leq 2N-1 \atop b_{ij}>0}\lambda[i]^{b_{ij}}+\prod_{1\leq i\leq 2N-1 \atop b_{ij}<0}\lambda[i]^{-b_{ij}}
\end{equation}
Which is a special case of the defining relations for a cluster algebra\cite{Fomin1,Fomin3,Williams}. 

With the Ryu-Takayanagi conjecture we can give the exchange relation for von-Neumann entropies as well. Using \eqref{eq:entropy_lambda} one can show that the following holds:

\begin{equation}\label{eq:entropy_recursion}
    S[j]+S'[j]=\frac{1}{2}\sum_{i=1}^{2N-1}|b_{ij}|S[i]+\frac{c}{3}\log{2\cosh{\frac{3}{2c}}\sum_{i=1}^{2N-1}b_{ij}S[i]}
\end{equation}
Notice that this formula is also true for the vacuum case, where we have shown\cite{LB} that the lambda lengths in that case determine an $A_{N-3}$ cluster algebra. The differences are only in the $B$ matrices constructed for the corresponding triangulations and in the number of independent entanglement entropies for a given number of CFT subsystems. Hence we obtained the result that in the case of the vacuum (dual to pure $AdS_3$) and the thermal state (dual to the BTZ black hole in the HTL), the CFT entanglement structures are encoded in $B$ matrices of cluster algebras. In both cases the \eqref{eq:entropy_recursion} recursion relation gives us an effective way to determine all of the $\mathcal{O}(N^2)$ entanglement entropies knowing only $\mathcal{O}(N)$ of such quantities.
 
\section{Kinematic space and $Y$-systems}

In this chapter we examine how our cluster algebraic structures manifest themselves in the space of directed geodesics, the so called kinematic space\cite{Czech1}. In Section III. we parametrized the geodesics on the Poincar\'e disk by $(B_1,B_2,M)$, used as coordinates in kinematic space. According to \eqref{eq:conserved}, the following relation holds for the parameters characterize geodesics: $B_1^2+B_2^2-M^2=1$. So we can think of the kinematic space as\cite{Czech1} a two dimensional $dS_2$ de Sitter space embedded in $\mathbb{R}^{2,1}$, endowed with the inner product

\begin{equation}
    ds^2_{\mathbb{K}}=dB_1^2+dB_2^2-dM^2
\end{equation}

\begin{figure}[t]
    \centering
    \includegraphics[width=0.9\columnwidth]{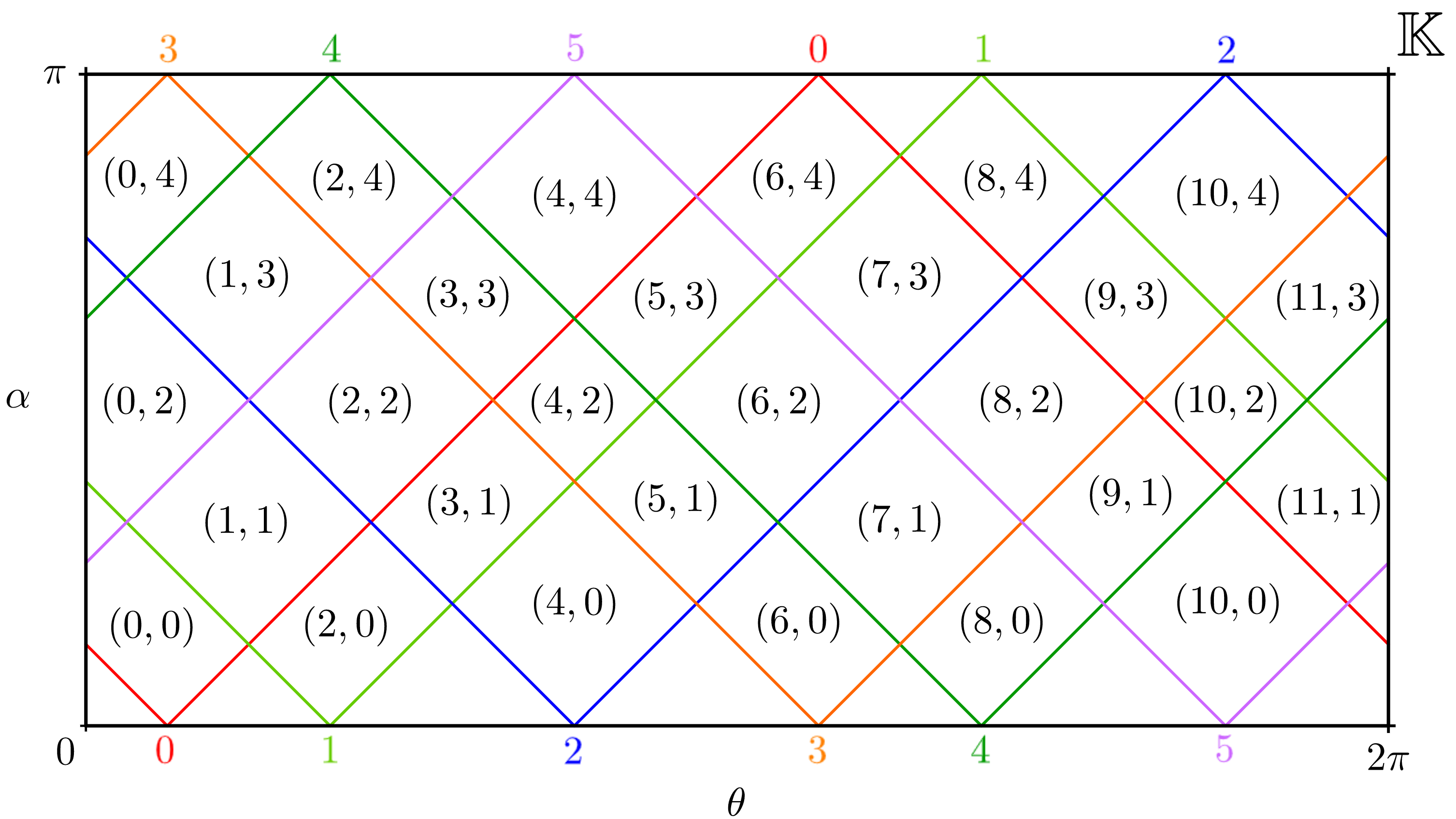}
    \caption{Kinematic space for the Poincar\'e disk. The colored point curves correspond to the vertices labeled by $(0,1,2,3,4,5)$ of a geodesic hexagon. The bounded rectangular domains labeled by the $(j,k)$ pairs are defined by \eqref{eq:jk}.}
    \label{fig:disk_kinematic}
\end{figure}

A more useful way to deal with the kinematic space is to use the $(\theta, \alpha)$ or the $(u,v)$ pairs from \eqref{eq:alfateta} and \eqref{eq:uv} as generalized coordinates. Applying the transformations of \eqref{eq:conserved} the induced metric is

\begin{equation}\label{eq:metric_kinematic}
d s_{\mathbb{K}}^{2}=\frac{d \theta^{2}-d \alpha^{2}}{\sin ^{2} \alpha}=\frac{dudv}{\sin ^{2} \frac{v-u}{2}}
\end{equation}
We can think of $(\theta,\alpha)$ as spacelike and timelike and $(u,v)$ as lightlike coordinates. The coordinate pairs $(\theta,\alpha)$ and $(\theta+\pi,\pi-\alpha)$ represent the same geodesic on $\mathbb{D}$ with different orientations. This means that the kinematic space of the whole Poincar\'e disk can be represented by the coordinate chart $(\theta,\alpha)\in[0,2\pi]\times[0,\pi]$, where $\theta\sim\theta+2\pi$ and a particular geodesic is represented by two points on the chart. On the other hand points of the Poincar\'e disk are represented by curves on the kinematic space. These are called point curves\cite{Czech1}. In the case of boundary points these are light-like straight lines (see \hyperref[fig:disk_kinematic]{FIG. 7.}).

If we are dealing with geodesic polygons on $\mathbb{D}$, we have got $N$ boundary points, giving rise to $N$ point curves on the kinematic space. They form a grid of the chart with rectangular domains. Let us choose two arbitrary $0\leq a<b\leq N-1\in\partial\mathbb{D}$ points such that

\begin{align}\label{eq:jk}
    a\equiv\frac{j-k}{2},&&b\equiv\frac{j+k}{2}&&\mod N
\end{align}
Where $j=0,\dots,2N-1$, $k=0,\dots, N-2$ and $j+k\equiv0 \mod2$. This gives us a $(j,k)$ coordinate set for the kinematic space tiles (see \hyperref[fig:disk_kinematic]{FIG. 7.}). The area of these tiles can be calculated in the $(u,v)$ representation

\begin{equation}\label{eq:area}
\begin{aligned}
    T_{j,k}&\equiv\int_{\vartheta_b}^{\vartheta_{b+1}} \int_{\vartheta_{a-1}}^{\vartheta_a} \frac{du\wedge dv}{4 \sin ^{2} \frac{v-u}{2}}=\\
    &=\log\frac{\sin \left(\frac{\vartheta_b-\vartheta_{a-1}
}{2}\right) \sin \left(\frac{\vartheta_{b+1}-\vartheta_a}{2}\right)}{\sin \left(\frac{\vartheta_b-\vartheta_a}{2}\right)  \sin \left(\frac{\vartheta_{b+1}-\vartheta_{a-1}
}{2}\right)}
\end{aligned}
\end{equation}
This cross ratio can also be expressed in terms of lambda lengths

\begin{equation}
    T_{j,k}=\log\frac{\lambda(a-1b)\lambda(ab+1)}{\lambda(ab)\lambda(a-1b+1)}
\end{equation}
Notice, that the area of the $k=0$ and $k=N-2$ domains are divergent.

The area form associated to the metric of \eqref{eq:metric_kinematic} is related to the Crofton form on kinematic space\cite{Czech1}

\begin{equation}
    \omega=\frac{\partial^2 S(u,v)}{\partial u\partial v}du\wedge dv =\frac{c}{12}\frac{du\wedge dv}{ \sin ^{2} \frac{v-u}{2}}
\end{equation}
Using this relation and comparing equations \eqref{eq:mutual} and \eqref{eq:area} for the vacuum state dual to pure $AdS_3$ one can relate every inner tile a conditional mutual information\cite{Czech1,LB}, namely

\begin{equation}
\begin{aligned}
    I_{j,k}&=I(a-1a,bb+1|ab)=\\
    &=S(a-1b)+S(ab+1)-S(ab)-S(a-1b+1)=\\
    &=\frac{c}{3}T_{j,k}
\end{aligned}
\end{equation}
The $k=0$ and $k=N-2$ tiles with divergent areas can be associated to mutual informations of the form $I(A,B)=S(A)+S(B)-S(AB)$ or labeled by the pointcurves $I(ab,cd)=S(a-1a)+S(bb+1)-S(a-1b+1)$, where $a=b$.

\begin{figure}[t]
    \centering
    \includegraphics[scale=0.4]{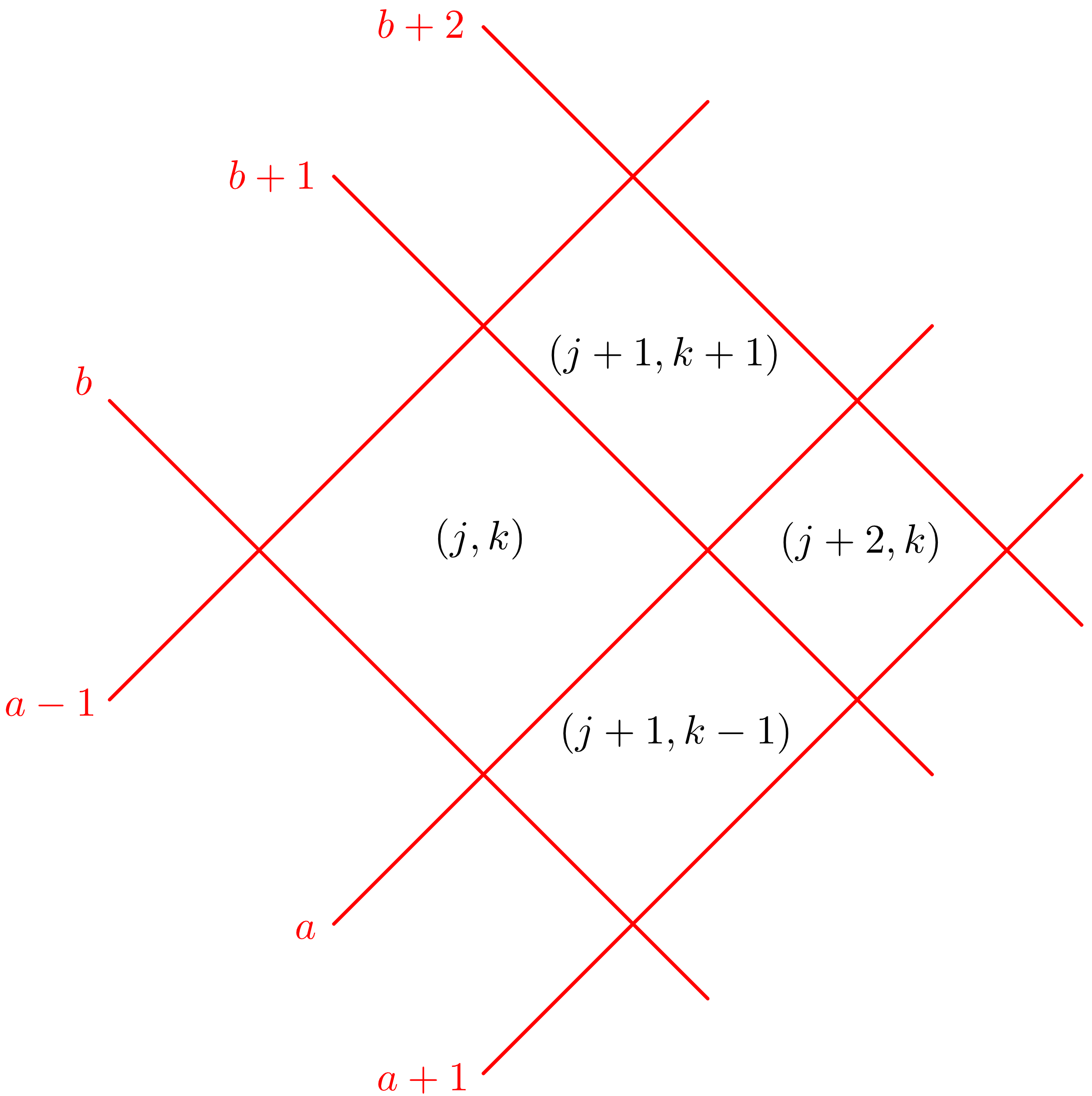}
    \caption{Four neighbouring tiles of the kinematic space in case of a geodesic $N$-gon. The tiles are labeled by $(j,k)$, $(j+1,k-1)$, $(j+1,k+1)$ and $(j,k+2)$. These regions are bounded by the point curves correspond to vertices $a-1$, $a$, $a+1$, $b$, $b+1$ and $b+2$ of the $N$-gon.}
    \label{fig:ysystem}
\end{figure}

Now consider four neighbouring tiles labeled by $(j,k)$, $(j+1,k-1)$, $(j+1,k+1)$ and $(j,k+2)$ (see \hyperref[fig:ysystem]{FIG. 8.}). These domains are bounded by the point curves corresponding to the $a-1$, $a$, $a+1$, $b$, $b+1$ and $b+2$ vertices. The areas of the four tiles:

\begin{subequations}
\begin{align}
    T_{j,k}&=\log\frac{\lambda(a-1b)\lambda(ab+1)}{\lambda(a-1b+1)\lambda(ab)},\\
    T_{j+1,k-1}&=\log\frac{\lambda(ab)\lambda(a+1b+1)}{\lambda(ab+1)\lambda(a+1b)},\\
    T_{j+1,k+1}&=\log\frac{\lambda(a-1b+1)\lambda(ab+2)}{\lambda(a-1b+2)\lambda(ab+1)},\\
    T_{j,k+2}&=\log\frac{\lambda(ab+1)\lambda(a+1b+2)}{\lambda(a+1b+1)\lambda(ab+2)}
\end{align}    
\end{subequations}
Let us introduce the following quantity for each domain:

\begin{equation}\label{eq:Y}
    Y_{j,k}=\frac{1}{e^{T_{j,k}}-1}=\frac{1}{e^{\frac{3}{c}I_{j,k}}-1}
\end{equation}
Using the definition of the lambda length, and trigonometric identities one can show the following relation

\begin{equation}\label{eq:recursion2}
    Y_{j,k}Y_{j+2,k}=(1+Y_{j+1,k-1})(1+Y_{j+1,k+1})
\end{equation}
Or by changing the label $j\rightarrow j-1$ (now $j+k\equiv1 \mod2$):
\begin{equation}\label{eq:recursion}
    Y_{j-1,k}Y_{j+1,k}=(1+Y_{j,k-1})(1+Y_{j,k+1})
\end{equation}
With the boundary conditions $Y_{j,0}=Y_{j,N-2}=0$.

As we have shown in our previous paper\cite{LB} in the case of the vacuum/pure $AdS_3$ duality these relations define a Zamolodchikov $Y$-system\cite{Zamo,FrenkelSzenes}.

\begin{figure*}[t]
    \hspace{0.2cm}
    \subfloat[]{
        \includegraphics[width=0.9\columnwidth]{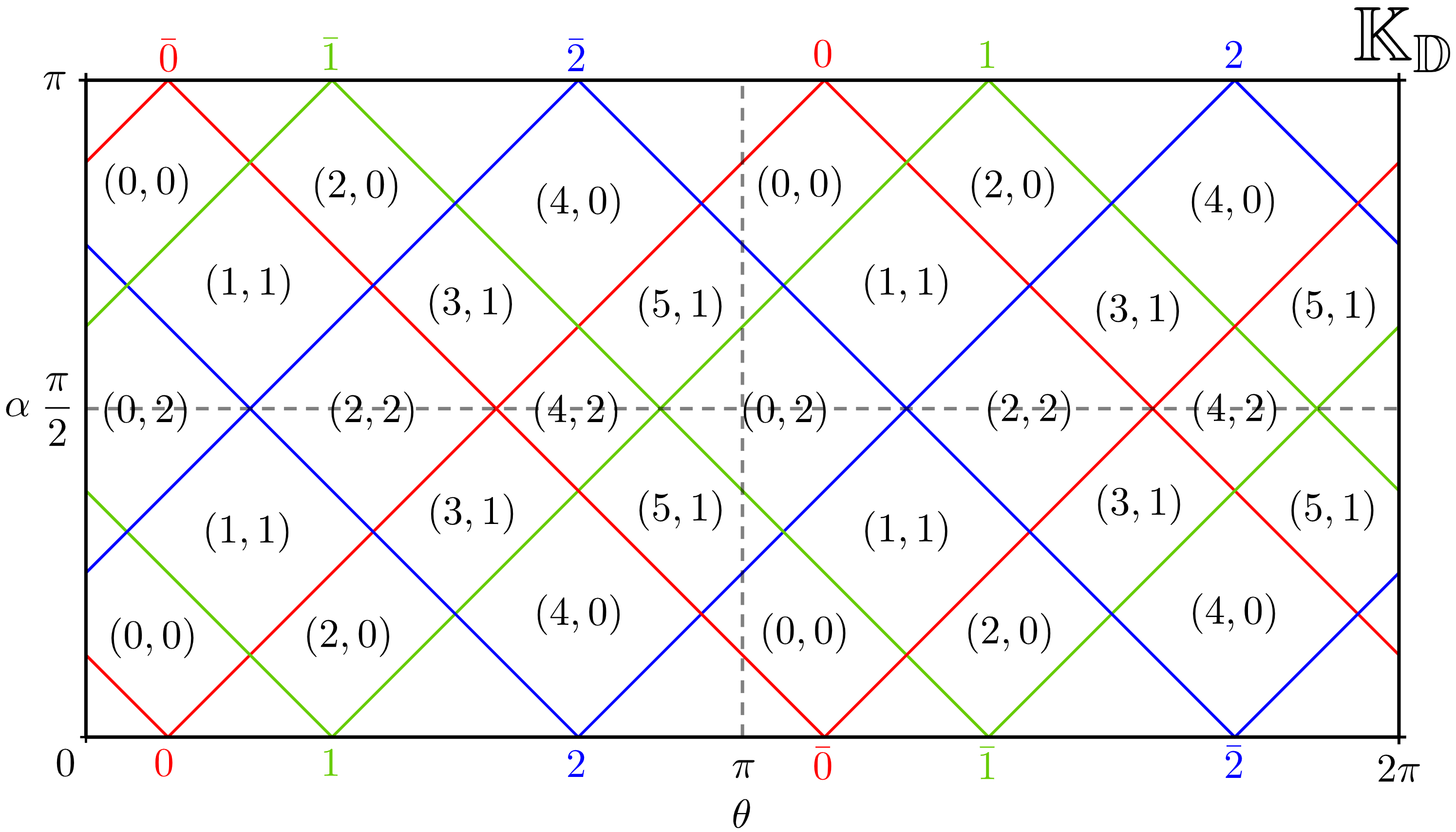}}
    \hfill
    \subfloat[]{
        \includegraphics[width=0.95\columnwidth]{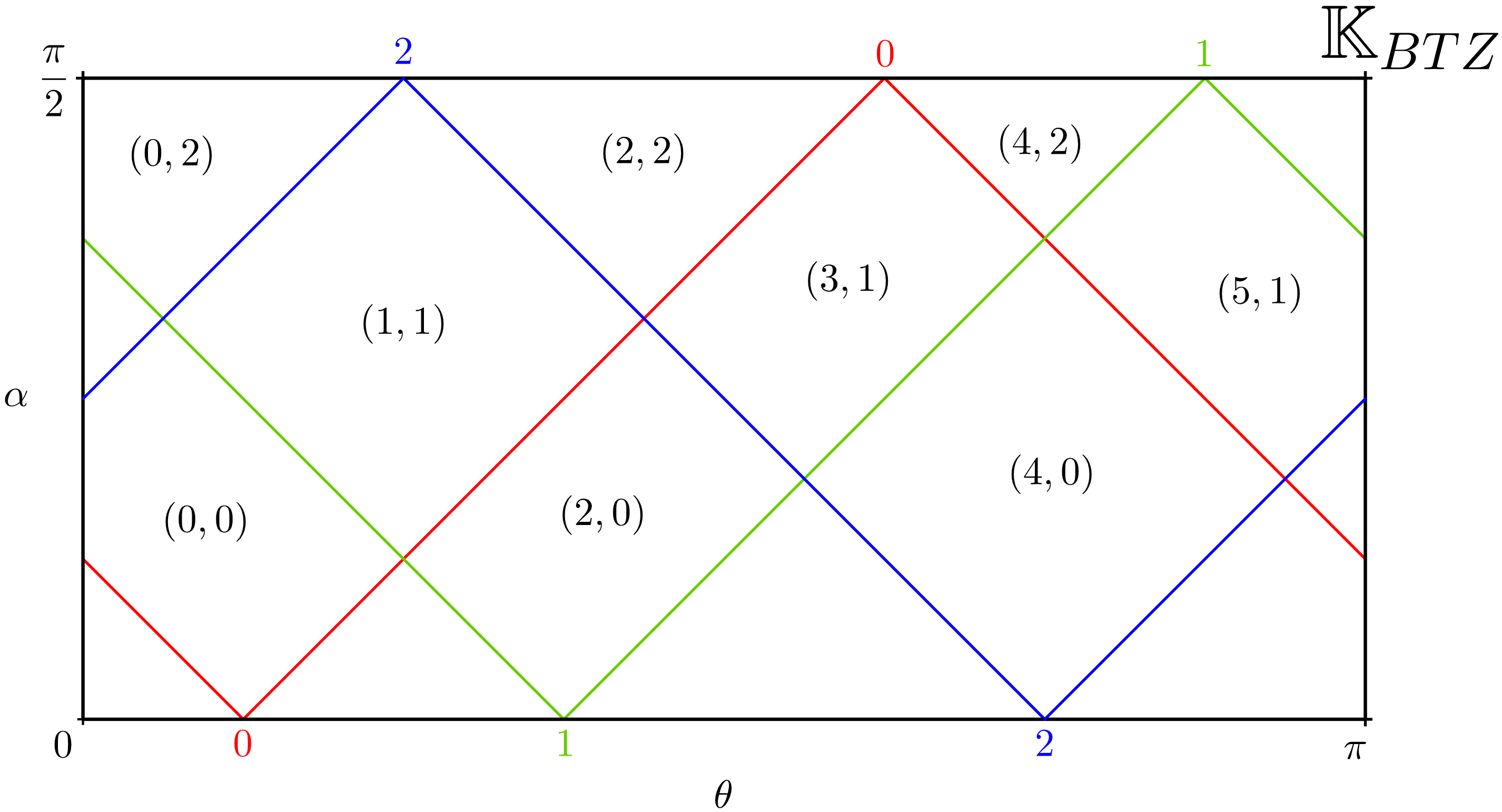}}
    \hspace{0.3cm}
    \caption{(a) The kinematic space of a Poincar\'e disk hexagon with vertices labeled by $a,b,c,\bar{a},\bar{b},\bar{c}$ such that $a,b,c$ are centrally symmetric to $\bar{a},\bar{b},\bar{c}$. In this special case kinematic space is build up from four identical domains, bordered by dashed grey lines. In the bottom left segment we labeled the tiles by the $(j,k)$ pairs defined by \eqref{eq:jk}. In the other three segments we labeled the tiles respectively. One can look at this space as the kinematic space of a $2N$-gon equivalent to a BTZ black hole $N$-gon. (b) A fundamental domain that fully represents the BTZ black hole $N$-gon. Notice that the uppermost triangles, now have finite areas.}
    \label{fig:kinematic_BTZ}
\end{figure*}

Based on this result a natural question to be asked is the following. What is the form of the $Y$-system for the thermal state/BTZ black hole duality?
In order to answer this question we turn to the kinematic space representation of BTZ geodesic polygons. Again we deal with the triangulations of the BTZ $N$-gon using a $2N$-gon on the Poincar\'e disk. This means that the kinematic space of the Poincar\'e disk (from now on denoted by $\mathbb{K}_{\mathbb{D}}$) with $2N$ point curves fully represent a BTZ $N$-gon triangulation. In this picture now we have centrally symmetric pairs of vertices and geodesics on $\mathbb{D}$, and the real BTZ black hole space is covered by just a half of the disk. As a result of this as a BTZ representative one can identify four identical fundamental domains on $\mathbb{K}_{\mathbb{D}}$, and one of them will completely describe the BTZ $N$-gon. Let us denote an arbitrary copy from these four domains by $\mathbb{K}_{BTZ}$. We can say that $\mathbb{K}_{BTZ}$ is the $[0,\pi]\times[0,\pi/2]$ quarter of the $[0,2\pi]\times[0,\pi]$ chart. For $\mathbb{K}_{\mathbb{D}}:\theta\sim\theta+2\pi$, but it is made up by two identical copies along $\theta$, and for $\mathbb{K}_{BTZ}:\theta\sim\theta+\pi$. The kinematic space of the disk representation is in \hyperref[fig:kinematic_BTZ]{FIG. 9. (a)}. The $\theta=\pi$ and $\alpha=\pi/2$ lines cut the kinematic space into four identical domains, and one of them fully represents the BTZ black hole. This domain is shown in \hyperref[fig:kinematic_BTZ]{FIG. 9. (b)}.

We can do the $(j,k)$ labeling for $\mathbb{K}_{\mathbb{D}}$ as before. But now, there are four identical fundamental domains, so each tile with a given area is included in the kinematic space four times. So we can label the tiles of one fundamental domain $\mathbb{K}_{BTZ}$ by the previous rules, and copy the labeling to the other three domains respectively to label $\mathbb{K}_{\mathbb{D}}$. We can choose the range of coordinates to be $j=0,1,\dots,2N+1$ and $k=0,1,\dots,N-1$, $j+k\equiv0\mod\,2$ to cover all the different tiles in $\mathbb{K}_{BTZ}$. The labeling for the disk representation is shown in \hyperref[fig:kinematic_BTZ]{FIG. 9. (a)} and for the BTZ representation is in \hyperref[fig:kinematic_BTZ]{FIG. 9. (b)}.

Let see what entanglement quantities are encoded in these tiles. The areas of $k=0$ domains are proportional to the divergent mutual informations, namely:
\begin{equation}
\begin{aligned}
    \frac{c}{3}T_{j,0}&=I[a-1a,bb+1]=\\
    &=S[a-1,a]+S[b,b+1]-S[a-1,b+1]   
\end{aligned}
\end{equation}
where $a=b$. Notice that we are using the square bracket notation since we are working in $\mathbb{K}_{BTZ}$. For $k\leq N-2$ we get conditional mutual informations of the form

\begin{equation}
\begin{aligned}
    \frac{c}{3}T_{j,k}&=I_{j,k}=\\
    &=I[a-1a,bb+1|ab]=\\
    &=S[a-1b]+S[ab+1]-S[ab]-S[a-1b+1]
\end{aligned}
\end{equation}
Let denote the areas of $k=N-1$ tiles (e.g. see \hyperref[fig:kinematic_BTZ]{Figure 8 (b) topmost triangles}) by $T_{j,N-1}$. Now we go back to $\mathbb{K}_{\mathbb{D}}$. In this picture the $k=N-1$ tiles (e.g. see \hyperref[fig:kinematic_BTZ]{Figure 8 (a) square in the middle strip}) have got areas $2\cdot T_{j,N-1}$. These squares are bounded by pointcurves $a,\bar{a},b,\bar{b}$, where

\begin{align}\label{eq:jk_BTZ}
    a\equiv\frac{j-k}{2},&&b\equiv\frac{j+k}{2},&&\bar{b}\equiv a-1,&&\bar{a} \equiv b+1&&\mod 2N
\end{align}
($a$ and $b$ can denote pointcurve with bar as well). So the areas can be expressed by entanglement entropies in the following way

\begin{equation}
    2\cdot\frac{c}{3}T_{j,N-1}=
I(ab,\overline{a}\overline{b}|b\overline{a})=
S(a\bar{a})+S(b\bar{b})-S(a\bar{b})-S(b\bar{a})
\end{equation}
where $I(ab,\overline{a}\overline{b}|b\overline{a})$ is the conditional mutual information we have calculated in \eqref{eq:later}.
Notice, that $S(a\bar{a})$ and $S(b,\bar{b})$ both gives the $S[\partial BTZ]$ von-Neumann entropy (which is nonzero since the thermal state is a mixed state) of the whole BTZ boundary and $S(a\bar{b})=S(b\bar{a})=S[ba]$. Now one can see that

\begin{equation}
\begin{aligned}
    S(a\bar{a})-S(a\bar{b})&=S(b\bar{b})-S(b\bar{a})=\\
    &=S[\partial BTZ]-S[ab]=\\
    &=S[ba|ab]
\end{aligned}
\end{equation}
Where $ba$ and $ab$ now represent BTZ boundary intervals and $S[ba|ab]=S[\partial BTZ]-S[ab]$ is a conditional entropy. Interestingly in this special case it is just the half of the conditional mutual information $I(ab,\overline{a}\overline{b}|b\overline{a})$ that we calculated in \eqref{eq:later}. If we return to the BTZ kinematic space one can also write

\begin{equation}
    \frac{c}{3}T_{j,N-1}=S[ba|ab]=\frac{1}{2}I(ab,\overline{a}\overline{b}|b\overline{a})
\end{equation}
Summarizing what we have so far

\begin{equation}
T_{j,k}=\frac{3}{c}\cdot\left\{\begin{array}{ll}
I[a-1a,bb+1], & \text { if } k=0 \\
I[a-1a,bb+1|ab], & \text { if } 0< k <N-1\\
S[ba,ab],& \text { if } k=N-1
\end{array}\right.
\end{equation}
Where 

\begin{align}
    a\equiv\frac{j-k}{2},&&b\equiv\frac{j+k}{2}&&\mod N
\end{align}

Now we want to derive an Y-system for the high-temperature BTZ case. Similarly to \eqref{eq:Y}, we can introduce the following quantities

\begin{equation}
Y_{j,k}=\left\{\begin{array}{ll}
\frac{1}{e^{T_{j,k}}-1}, & \text { if } 0\leq k <N-1 \\
\frac{1}{e^{2T_{j,k}}-1}, & \text { if } k=N-1
\end{array}\right.
\end{equation}
Where $T_{j,k}$ are the areas of tiles in $\mathbb{K}_{BTZ}$. With the $\mathbb{K}_\mathbb{D}$ representation one can give two types of relations between different tiles. One for the inner tiles of a fundamental domain

\begin{equation}
    Y_{j-1,k}Y_{j+1,k}=(1+Y_{j,k-1})(1+Y_{j,k+1})
\end{equation}
Where $k\leq N-2$. And one for the topmost tiles, shared by two fundamental domains:
\begin{equation}
    Y_{j-1,N-1}Y_{j+1,N-1}=(1+Y_{j,N-2})^2
\end{equation}
These relations determine a Zamolodchikov $Y$-system of $C_{N-1}$ type

\begin{equation}
Y_{j-1,k} Y_{j+1,k}=\prod_{i \neq k}\left(Y_{j,i}+1\right)^{-a_{ki}}
\end{equation}
Where the boundary conditions are now
\begin{align}
    Y_{j,0}=0,&&Y_{j,N-1}=\frac{1}{e^{2T_{j,N-1}}-1}=\frac{1}{e^{2S[ba,ab]}-1}
\end{align}
Here $a_{ki}$ is the Cartan matrix of the $C_{N-1}$ Dynkin diagram. In general its solutions are periodic in the variable $j$ with period $4N$ which is inherited from the $\theta\sim\theta+2\pi$ periodicity of kinematic space. However, since now the $2N$-gon being centrally symmetric, the period in our case is $2N$.
Notice that by virtue of \eqref{eq:later} the boundary conditions for $Y_{j,N-1}$ are featuring the Bekenstein-Hawking entropy of the BTZ black hole

\begin{equation}
\label{eq:BTZblack}
S_{BH}=\frac{2\pi r_+}{4G}
\end{equation}
where
$2\pi r_+$ is the area of the black hole.

\section{Conclusions}
Within the framework of the $AdS_3/CFT_2$ correspondence in this paper we investigated how entangled quantum states of the boundary are encoded into the classical geometric structure of the bulk. In our previous work\cite{LB} we have shown that the entanglement patterns of the CFT vacuum are encoded into the geometry of pure $AdS_3$ by the algebraic structure of a cluster algebra. For a partitioning of the boundary into $N$ regions this algebra turned out to be of $A_{N-3}$ type.
After this observation the natural question to be asked was the one: are there any other interesting cases where we again find this particular type of encoding via cluster algebras?
In this work we have shown that the answer to this question is yes. We have shown that the entanglement patterns of a thermal state of the boundary are encoded into the high temperature limit of the static BTZ geometry via another type of a cluster algebra. For a similar partitioning of the boundary to $N$ regions this is of type $C_{N-1}$.   

One can study this encoding phenomenon in the bulk or in kinematic space.
In the bulk case the cluster algebraic structure manifests itself in algebraic relations between the regularized (lambda) lengths of geodesics. On the other hand in the kinematic space description this structure is captured by relations between areas of causal diamonds with respect to the Crofton form. For our examples studied so far the kinematic space version of this encoding has given rise to Zamolodchikov Y-systems of type $A_{N-3}$ (vacuum) and $C_{N-1}$ (thermal state).
We also observed that in the $C_{N-1}$ case the boundary conditions for the $Y$ system display in the explicit form of the $Y_{j,N-1}$ quantities the Bekenstein-Hawking entropy of the BTZ black hole.

We note that interestingly in the language of cluster algebras\cite{Williams} in the bulk representation the encoding manifests itself via cluster dynamics of flips , and in the kinematic space representation by the so called coefficient dynamics of flips. In physical terms cluster dynamics is the one based on mutation between possible partitions of the boundary 
captured by regularized entanglement entropies. On the other hand coefficient dynamics is the one based on similar mutations encapsulating changes in regularization independent conditional mutual informations.

The advantage of studying this encoding phenomenon with the help of algebraic structures is particularly transparent in kinematic space. 
Here one can investigate the dynamics of cross ratios which are gauge invariant quantities, meaning that they are independent of the regularization prescription. Moreover, one also has the physical interpretation of cross ratios as conditional mutual informations subject to strong subadditivity. This constraint  gives rise to further interesting connections with the topic of positive geometry which is an important ingredient of recent studies on scattering amplitudes\cite{Nima,Assoc}.

Our investigations also revealed an interesting connection between quantum entanglement on the boundary and cluster polytopes. These cluster polytopes\cite{Nima,Nima1} are playing a very important role in the rapidly evolving research field on scattering amplitudes. Such research studies culminated in the appearance of the amplituhedron a polytopal object geometrizing the factorization properties of scattering amplitudes\cite{Ampli}. Now in this new context we have found that for an $N$-fold partitioning of the boundary the associahedron ${\mathcal A}_{N-3}$ geometrizes entaglement information of the vacuum and the cyclohedron ${\mathcal C}_{N-1}$ is doing the same for the thermal state.
Since these objects are encoding holographic entanglement information in a polytopal manner, they can be regarded as some sort of holographic entanglement polytopes\cite{Assoc}.
However, this term
should be handled with care not be confused with the
existing topic of entanglement polytopes in the quantum
information\cite{Borland,Klyachko,Saw} and in the holographic context\cite{Stoica,Hub1,Hub2,Hub3}.
In any case the associahedron ${\mathcal A}_{N-3}$ for example can be visualized as a polytope existing in a $N-3$ dimensional Euclidean space. This space is spanned by the {\it regularized} entanglement entropies associated to the diagonals of the quadrangles arising from a particular triangulation of the bulk. Then the associahedron is cut out from this space by the positivity constraints dictated by strong subadditivity\cite{Assoc}.
Clearly this polytopal type of encoding of holographic entanglement information should be further investigated.

Finally we note that cluster algebras originally appeared implicitely in Teichm\"uller thory of Riemann surfaces. In this context one should bear in mind that one can associate a cluster algebra to any bordered surface with marked points\cite{Williams}. For example this construction specializes to our type $A_{N-3}$ case dual to the $CFT_2$ vacuum. In this special case the surface is just a disk with $N$ marked points. Since in the $AdS_3/CFT_2$ context multiboundary wormhole solutions are naturally showing up as ones featuring such surfaces\cite{Skenderis} one expects that the examples investigated in our paper provide just the simplest ones based on a generic construction.
This conjectured encoding\cite{Levay} of quantum states in a holographic manner via cluster algebras and their associated cluster polytopes is certainly an interesting possibility worth exploring in the future.

\bigskip
\section{Acknowledgement}

This work was supported by the National Research Development and Innovation Office of Hungary within the Quantum Technology National Excellence Program (Project No. 2017-1.2.1-NKP-2017-0001).
Supported by the ÚNKP-20-1 New National Excellence Program of the Ministry for Innovation and Technology from the source of National Research, Development and Innovation Fund.

\end{document}